\documentclass{nature}
\pdfoutput=1
\usepackage{hyperref}
\hypersetup{colorlinks,citecolor=black,filecolor=black,linkcolor=black,urlcolor=black}
\usepackage{url}
\usepackage{graphicx}
\usepackage{amssymb}
\usepackage{xcolor}
\usepackage{siunitx}
\usepackage{booktabs}
\usepackage{caption}
\usepackage{aas_macros}

\newcounter{firstbib}

\title{Teraelectronvolt emission from the $\gamma$-ray burst GRB~190114C}

\author{
MAGIC Collaboration:
V.~A.~Acciari$^{1}$,
S.~Ansoldi$^{2,22}$,
L.~A.~Antonelli$^{3}$,
A.~Arbet Engels$^{4}$,
D.~Baack$^{5}$,
A.~Babi\'c$^{6}$,
B.~Banerjee$^{7}$,
U.~Barres de Almeida$^{8}$,
J.~A.~Barrio$^{9}$,
J.~Becerra Gonz\'alez$^{1}$
% ...
}

\begin{document}

\maketitle


\begin{affiliations}
\item {Inst. de Astrof\'isica de Canarias, E-38200 La Laguna, and Universidad de La Laguna, Dpto. Astrof\'isica, E-38206 La Laguna, Tenerife, Spain} 
\item {Universit\`a di Udine, and INFN Trieste, I-33100 Udine, Italy} 
\item {National Institute for Astrophysics (INAF), I-00136 Rome, Italy} 
\item {ETH Zurich, CH-8093 Zurich, Switzerland} 
\item {Technische Universit\"at Dortmund, D-44221 Dortmund, Germany} 
\item {Croatian Consortium: University of Rijeka, Department of Physics, 51000 Rijeka; University of Split - FESB, 21000 Split; University of Zagreb - FER, 10000 Zagreb; University of Osijek, 31000 Osijek; Rudjer Boskovic Institute, 10000 Zagreb, Croatia} 
\item {Saha Institute of Nuclear Physics, HBNI, 1/AF Bidhannagar, Salt Lake, Sector-1, Kolkata 700064, India} 
\item {Centro Brasileiro de Pesquisas F\'isicas (CBPF), 22290-180 URCA, Rio de Janeiro (RJ), Brasil} 
\item {IPARCOS Institute and EMFTEL Department, Universidad Complutense de Madrid, E-28040 Madrid, Spain} 
\item {University of \L\'od\'z, Department of Astrophysics, PL-90236 \L\'od\'z, Poland} 
\item {Universit\`a di Siena and INFN Pisa, I-53100 Siena, Italy} 
\item {Deutsches Elektronen-Synchrotron (DESY), D-15738 Zeuthen, Germany} 
\item {Istituto Nazionale Fisica Nucleare (INFN), 00044 Frascati (Roma) Italy} 
\item {Max-Planck-Institut f\"ur Physik, D-80805 M\"unchen, Germany} 
\item {Institut de F\'isica d'Altes Energies (IFAE), The Barcelona Institute of Science and Technology (BIST), E-08193 Bellaterra (Barcelona), Spain} 
\item {Universit\`a di Padova and INFN, I-35131 Padova, Italy} 
\item {Universit\`a di Pisa, and INFN Pisa, I-56126 Pisa, Italy} 
\item {Centro de Investigaciones Energ\'eticas, Medioambientales y Tecnol\'ogicas, E-28040 Madrid, Spain} 
\item {Universit\"at W\"urzburg, D-97074 W\"urzburg, Germany} 
\item {Finnish MAGIC Consortium: Finnish Centre of Astronomy with ESO (FINCA), University of Turku, FI-20014 Turku, Finland; Astronomy Research Unit, University of Oulu, FI-90014 Oulu, Finland} 
\item {Departament de F\'isica, and CERES-IEEC, Universitat Aut\`onoma de Barcelona, E-08193 Bellaterra, Spain} 
\item {Japanese MAGIC Consortium: ICRR, The University of Tokyo, 277-8582 Chiba, Japan; Department of Physics, Kyoto University, 606-8502 Kyoto, Japan; Tokai University, 259-1292 Kanagawa, Japan; RIKEN, 351-0198 Saitama, Japan} 
\item {The Armenian Consortium: ICRANet-Armenia at NAS RA, A. Alikhanyan National Laboratory} 
\item {Inst. for Nucl. Research and Nucl. Energy, Bulgarian Academy of Sciences, BG-1784 Sofia, Bulgaria} 
\item {Universitat de Barcelona, ICCUB, IEEC-UB, E-08028 Barcelona, Spain} 
\item {also at Dipartimento di Fisica, Universit\`a di Trieste, I-34127 Trieste, Italy}
\item {also at Port d'Informaci\'o Cient\'ifica (PIC) E-08193 Bellaterra (Barcelona) Spain}
\item {now at Laboratoire d'Annecy de Physique des Particules, Univ. Grenoble Alpes, Univ. Savoie Mont Blanc, CNRS, LAPP, 74000 Annecy, France}
\item {Istituto Nazionale Fisica Nucleare (INFN), I-34149 Trieste, Italy}
\item {Institute for Fundamental Physics of the Universe (IFPU), I-34151 Trieste, Italy}
\end{affiliations}

\begin{abstract}
Gamma-ray bursts (GRBs) of the long-duration class are the most luminous sources of electromagnetic radiation known in the Universe. They are generated by outflows of plasma ejected at near the speed of light by newly formed neutron stars or black holes of stellar mass at cosmological distances\cite{Gehrels&Meszaros2012,Kumar&Zhang2015}.
Prompt flashes of MeV gamma rays are followed by longer-lasting afterglow emission from radio waves to GeV gamma rays, due to synchrotron radiation by energetic electrons in accompanying shock waves\cite{Meszaros2002,Piran2004}.
Although emission of gamma rays at even higher, TeV energies by other radiation mechanisms had been theoretically predicted\cite{Meszarosetal2004,Fan&Piran2008,Inoueetal2013,Nava2018}, it had never been detected previously\cite{Inoueetal2013,Nava2018}.
Here we report the clear detection of GRB 190114C in the TeV band, achieved after many years of dedicated searches for TeV emission from GRBs.
Gamma rays in the energy range 0.2--1 TeV are observed from about 1 minute after the burst (at more than 50 standard deviations in the first 20 minutes).
This unambiguously reveals a new emission component in the afterglow of a GRB, whose power is comparable to that of the synchrotron component.
The observed similarity in the radiated power and temporal behaviour of the TeV and X-ray bands points to processes such as inverse Compton radiation as the mechanism of the TeV emission \cite{Meszarosetal1994,Zhang&Meszaros2001,Beniaminietal2015},
while processes such as synchrotron emission by ultrahigh-energy protons
\cite{Vietri1997,Boettcher&Dermer1998,Zhang&Meszaros2001}
are disfavoured due to their low radiative efficiency.
\end{abstract}

GRB\,190114C was first identified as a long-duration GRB by the BAT instrument onboard the Neil Gehrels Swift Observatory ({\it Swift})\cite{Gropp2019}
and the Gamma-ray Burst Monitor (GBM) instrument onboard the {\it  Fermi} satellite\cite{Hamburg2019}
on 14 January 2019, 20:57:03\, Universal Time (UT) (hereafter $T_0$).
Its duration in terms of $T_{90}$ (time interval containing 90\% of the total photon counts) was measured to be $\sim 116$\,s by {\it Fermi}/GBM \cite{Hamburg2019} and $\sim 362$\,s by {\it Swift}/BAT\cite{Krimmetal2019}.
Soon afterwards, reports followed on the detection of its afterglow emission at various wavebands from 1.3\,GHz up to 23\,GeV (Acciari et al., in preparation) and the measurement of its redshift $z = 0.4245 \pm 0.0005$\cite{Selsing2019,Castro-Tirado2019} (corresponding to cosmic distance).
The isotropic-equivalent energy of the emission at $\varepsilon=$10--1000 keV during $T_{90}$ observed by {\it Fermi}/GBM was $E_{\rm iso}\sim 3 \times 10^{53}$\,erg, implying that GRB\,190114C was fairly energetic, but not exceptionally so compared to previous events (Methods).

\begin{figure}[htb!]
\centering
\includegraphics[width=0.9\textwidth]{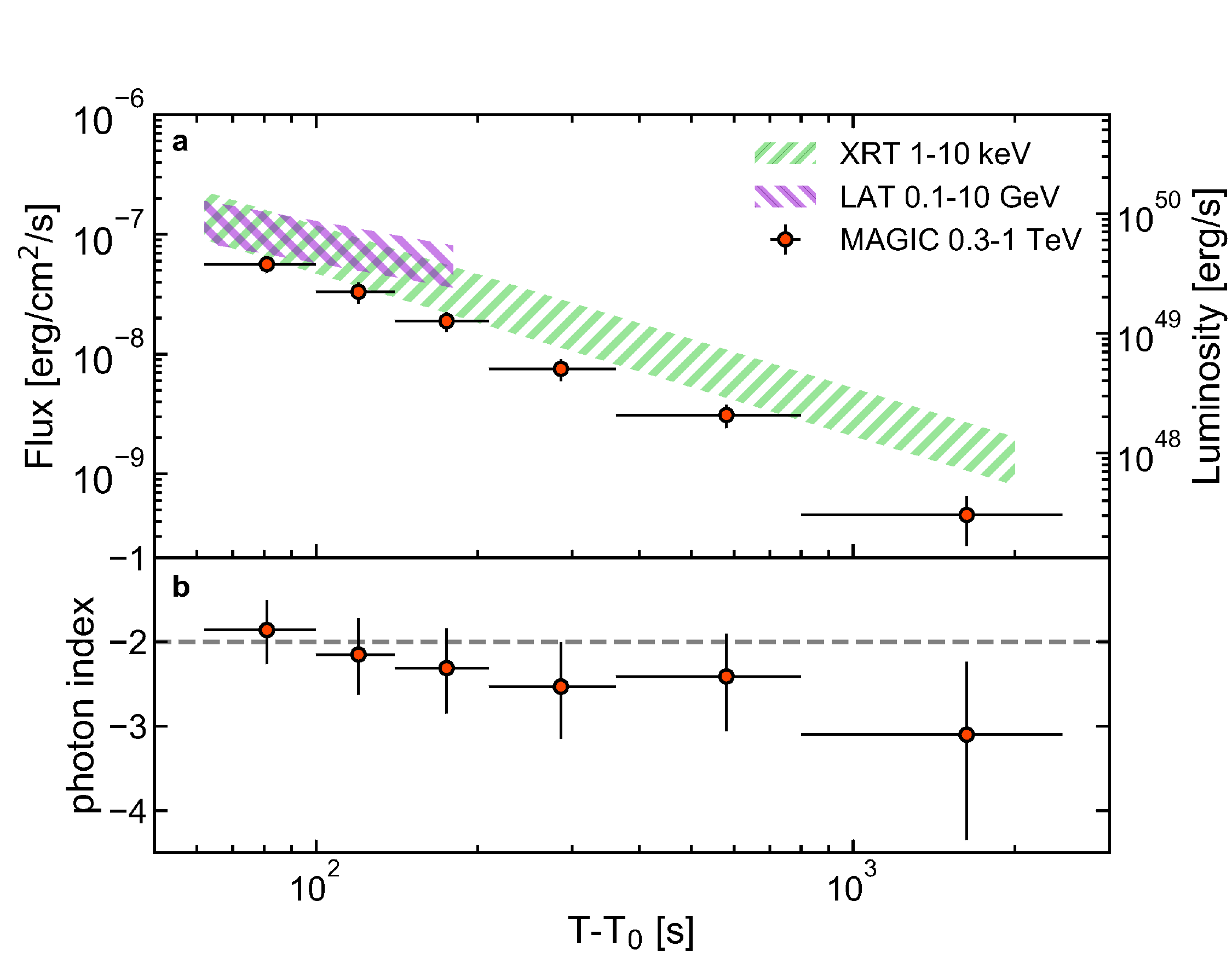}
\caption{
{\bf Light curves in the keV, GeV and TeV bands, and spectral evolution in the TeV band for GRB~190114C.}
{\bf a}, Light curves in units of energy flux (left axis) or apparent luminosity (right axis), for MAGIC at $0.3-1$\,TeV (red symbols), {\it Fermi}/LAT at $0.1-10$\,GeV (purple band) and {\it Swift}/XRT at $1-10$\,keV (green band).
For MAGIC, the intrinsic flux is shown, corrected for EBL attenuation\cite{Dominguezetal11} from the observed flux.
{\bf b}, Temporal evolution of the power-law photon index determined from time-resolved intrinsic spectra. The horizontal dashed line indicates the value -2. The errors shown in both panels are statistical only (1 standard deviation).}
\label{fig:full_MAGIC_lc}
\end{figure}

Triggered by the {\it Swift}/BAT alert, the Major Atmospheric Gamma Imaging Cherenkov (MAGIC) telescopes\cite{Aleksicetal2016a,Aleksicetal2016b}
observed GRB 190114C from $T_0+57$ seconds until $T_0+15912$ seconds (Extended Data Figure\,\ref{fig:full_MAGIC_lc}).
Gamma rays above 0.2\,TeV were detected with high significance from the beginning of the observations\cite{Mirzoyan2019}; in the first 20 minutes of data, the significance of the total gamma-ray signal is more than 50 standard deviations (Methods, Extended Data Figure\,\ref{fig:grb190114c_theta2_first_20min}).

For cosmologically distant objects such as GRBs, the observed gamma-ray spectra can be substantially modified due to attenuation by the extragalactic background light (EBL)\cite{Dwek&Krennrich2013}.
The EBL is the diffuse background of infrared, optical and ultraviolet radiation that permeates intergalactic space, constituting the emission from all galaxies in the Universe.
Gamma rays can be effectively absorbed during their propagation via photon-photon pair production interactions with low-energy photons of the EBL, which is more severe for higher photon energies and higher redshifts.
The gamma-ray spectrum that would be observed if the EBL was absent, referred to as the intrinsic spectrum, can be inferred from the observed spectrum by ``correcting'' for EBL attenuation, assuming a plausible model of the EBL\cite{Dominguezetal11}.

Emission from GRBs occurs in two stages that can partially overlap in time.
The ``prompt'' emission phase is characterised by a brief but intense flash of gamma rays, primarily at MeV energies. It exhibits irregular variability on timescales shorter than milliseconds, and lasts up to hundreds of seconds for long-duration GRBs.
These gamma rays are generated in the inner regions of collimated jets of plasma, which are ejected with ultra-relativistic velocities from highly magnetised neutron stars or black holes that form following the death of massive stars\cite{Kumar&Zhang2015}.
The ensuing ``afterglow'' phase is characterised by emission that spans a broader wavelength range and decays gradually over much longer timescales compared to the prompt emission.
This originates from shock waves caused by the interaction of the jet with the ambient gas (``external shocks''). Its evolution is typified by power-law decay in time due to the self-similar properties of the decelerating shock wave\cite{Meszaros2002,Piran2004}.
The afterglow emission of previously observed GRBs, from radio frequencies to GeV energies, is generally interpreted as synchrotron radiation from energetic electrons that are accelerated within magnetised plasma at the external shock\cite{Kumar&Zhang2015}.
Clues to whether the newly observed TeV emission is associated with the prompt or the afterglow phase are offered by the observed light curve (flux $F(t)$ as a function of time $t$).

Fig.~\ref{fig:full_MAGIC_lc} shows such a light curve for the EBL-corrected intrinsic flux in the energy range $\varepsilon=0.3-1$\,TeV (see also Extended Data Table\,\ref{tab:energy_flux_lc}).
It is well fit with a simple power-law function $F(t)\propto t^\beta$ with $\beta=-1.60\pm0.07$.
The flux evolves from $F(t)\sim5\times10^{-8}$\,erg\,cm$^{-2}$\,s$^{-1}$ at $t\sim T_0+$\,80\,s to $F(t)\sim6\times10^{-10}$\,erg\,cm$^{-2}$\,s$^{-1}$ at $t\gtrsim T_0+10^3$\,s, after which it falls below the sensitivity level and is undetectable.
There is no clear evidence for breaks or cutoffs in the light curve, nor irregular variability beyond the monotonic decay.
The light curves in the keV and GeV bands display behaviour similar to the TeV band, with somewhat shallower decay slope for the GeV band (Fig.~\ref{fig:full_MAGIC_lc}).
These properties indicate that most of the observed emission is associated with the afterglow phase, rather than the prompt phase that typically shows irregular variability.
Note that while the measured $T_{90}$ is as long as $\sim$360 sec, the keV-MeV emission does not exhibit clear temporal or spectral evidence for a prompt component after $\sim T_0+$\,25\,s \cite{Ravasioetal2019} (Methods).
Nevertheless, a sub-dominant contribution to the TeV emission from a prompt component at later times cannot be excluded.
The flux initially observed at $t\sim T_0+$\,80\,s corresponds to apparent isotropic-equivalent luminosity $L_{\rm iso}\sim3\times10^{49}$\,erg\,s$^{-1}$ at $\varepsilon=0.3-1$\,TeV, making this the most luminous source known at these energies.

The power radiated in the TeV band is comparable, within a factor of $\sim2$, to that in the soft X-ray and GeV bands, during the periods when simultaneous TeV-keV or TeV-GeV data are available (Fig.~\ref{fig:full_MAGIC_lc}).
The isotropic-equivalent energy radiated at $\varepsilon=0.3-1$\,TeV integrated over the time period between $T_0+62$ seconds and $T_0+2454$ seconds is $E_{\rm 0.3-1\,TeV} \sim 4\times10^{51}$\,erg. This is a lower limit to the total TeV-band output, as it does not account for data before $T_0+62$ seconds, nor potential emission at $\varepsilon >1$\,TeV.
The start of the power-law decay phase inferred from MeV-GeV data is $T_0+6\,s$ \cite{Wangetal2019,Ravasioetal2019}.
Assuming that the MAGIC light curve evolved as $F(t)\propto t^{-1.60}$ from this time, the TeV-band energy integrated between $T_0+6$ seconds and $T_0+2454$ seconds
is $E_{\rm 0.3-1\,TeV} \sim 2\times10^{52}$\,erg.
This would be $\sim$\,10\% of $E_{\rm iso}$ as measured by {\it Fermi}/GBM at $\varepsilon=$10--1000 keV.

Fig.~\ref{fig:full_MAGIC_lc} also shows the time evolution of the intrinsic spectral photon index $\alpha_{\rm int}$, determined by fitting the EBL-corrected, time-dependent differential photon spectrum with the power-law function $dF/d\varepsilon \propto \varepsilon^{\alpha_{\rm int}}$.
Considering the statistical and systematic errors (Methods), there is no significant evidence for spectral variability.
Throughout the observations, the data are consistent with $\alpha_{\rm int} \sim -2$, indicating that the radiated power is nearly equally distributed in $\varepsilon$ over this band.

\begin{figure}[htb!]
\centering
\includegraphics[width=0.9\textwidth]{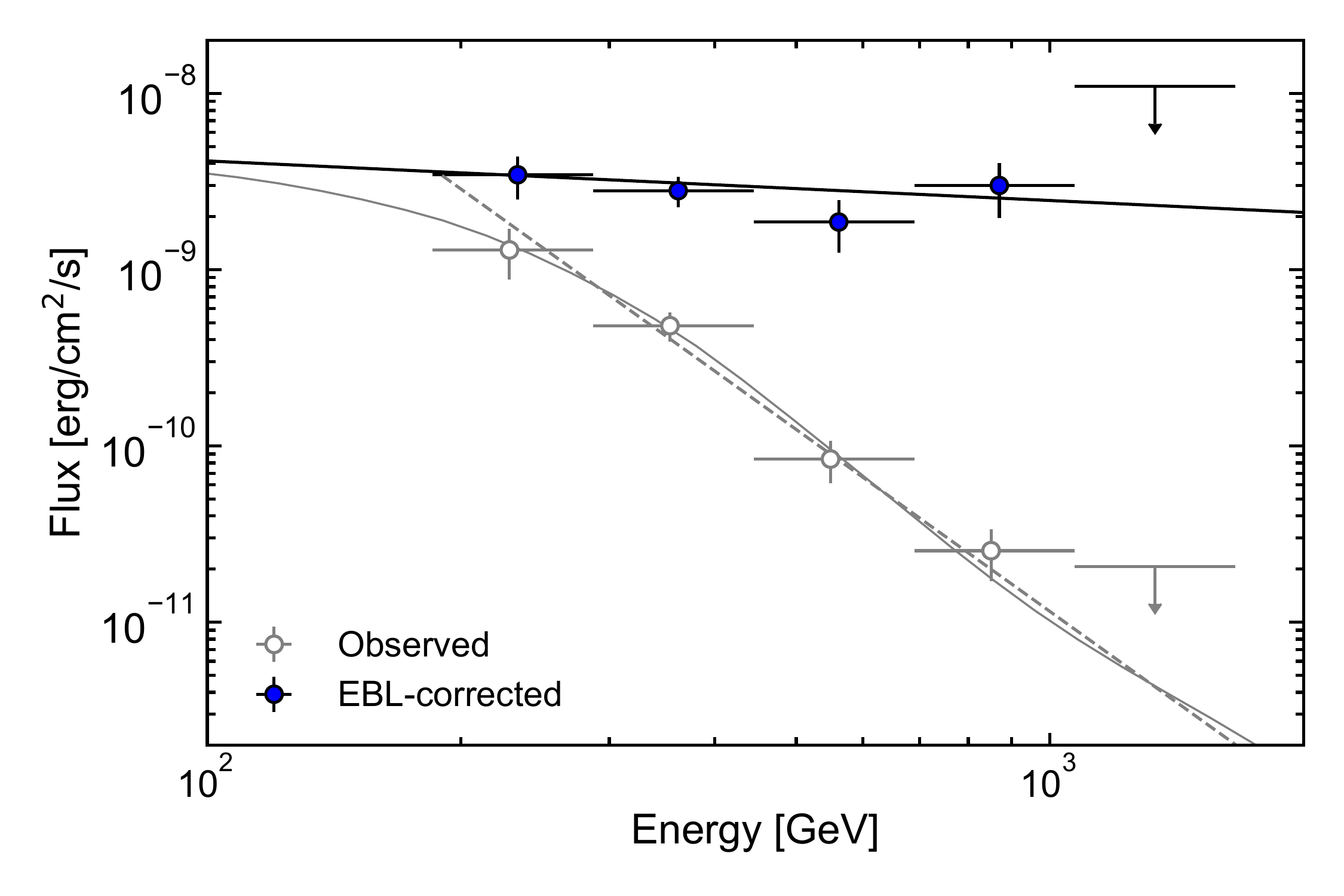}
\caption{
{\bf Spectrum above 0.2\,TeV averaged over the period between $T_0+$62\,s and $T_0+$2454\,s for GRB 190114C.}
Spectral energy distributions for the spectrum observed by MAGIC (grey open circles) and the intrinsic spectrum corrected for EBL attenuation\cite{Dominguezetal11} (blue filled circles).
The errors on the flux correspond to 1 standard deviation. The upper limits at 95\% confidence level are shown for the first non-significant bin at high energies.
Also shown is the best fit model for the intrinsic spectrum (black curve), when assuming a power-law function.
The grey solid curve for the observed spectrum is obtained by convolving this curve with the effect of EBL attenuation.
The grey dashed curve is the forward-folding fit to the observed spectrum with a power-law function (Methods).}
\label{fig:MAGIC_SED_timeinteg}
\end{figure}

Fig.~\ref{fig:MAGIC_SED_timeinteg} presents both the observed and the EBL-corrected intrinsic flux spectra above 0.2 TeV, averaged over ($T_0+$62\,s, $T_0+$2454\,s) when the GRB is detected by MAGIC. 
The observed spectrum can be fit in the energy range $0.2-1$\,TeV with a simple power-law with photon index $\alpha_{\rm obs} = -5.43 \pm 0.22$ (statistical error only),
one of the steepest spectra ever observed for a gamma-ray source.
It is remarkable that photons are observed at $\varepsilon \sim$\,1\,TeV (Extended Data Table\,\ref{tab:excess_counts}), despite the severe EBL attenuation expected at these energies (by a factor $\sim$ 300 based on plausible EBL models (Methods)).
Assuming a particular EBL model\cite{Dominguezetal11},
the intrinsic spectrum is well described as a power-law with $\alpha_{\rm int}=-2.22^{+0.23}_{-0.25}$ (statistical error only),
extending beyond 1~TeV at 95\% confidence level with no evidence for a spectral break or cutoff (Methods).
Adopting other EBL models leads to only small differences in $\alpha_{\rm int}$, compatible within the uncertainties (Methods).
Consistency with $\alpha_{\rm int} \sim -2$ implies roughly equal power radiated over $0.2-1$\,TeV and possibly beyond, strengthening the inference that there is significant energy output at TeV energies.

\begin{figure}[htb!]
\centering
\includegraphics[width=0.9\textwidth]{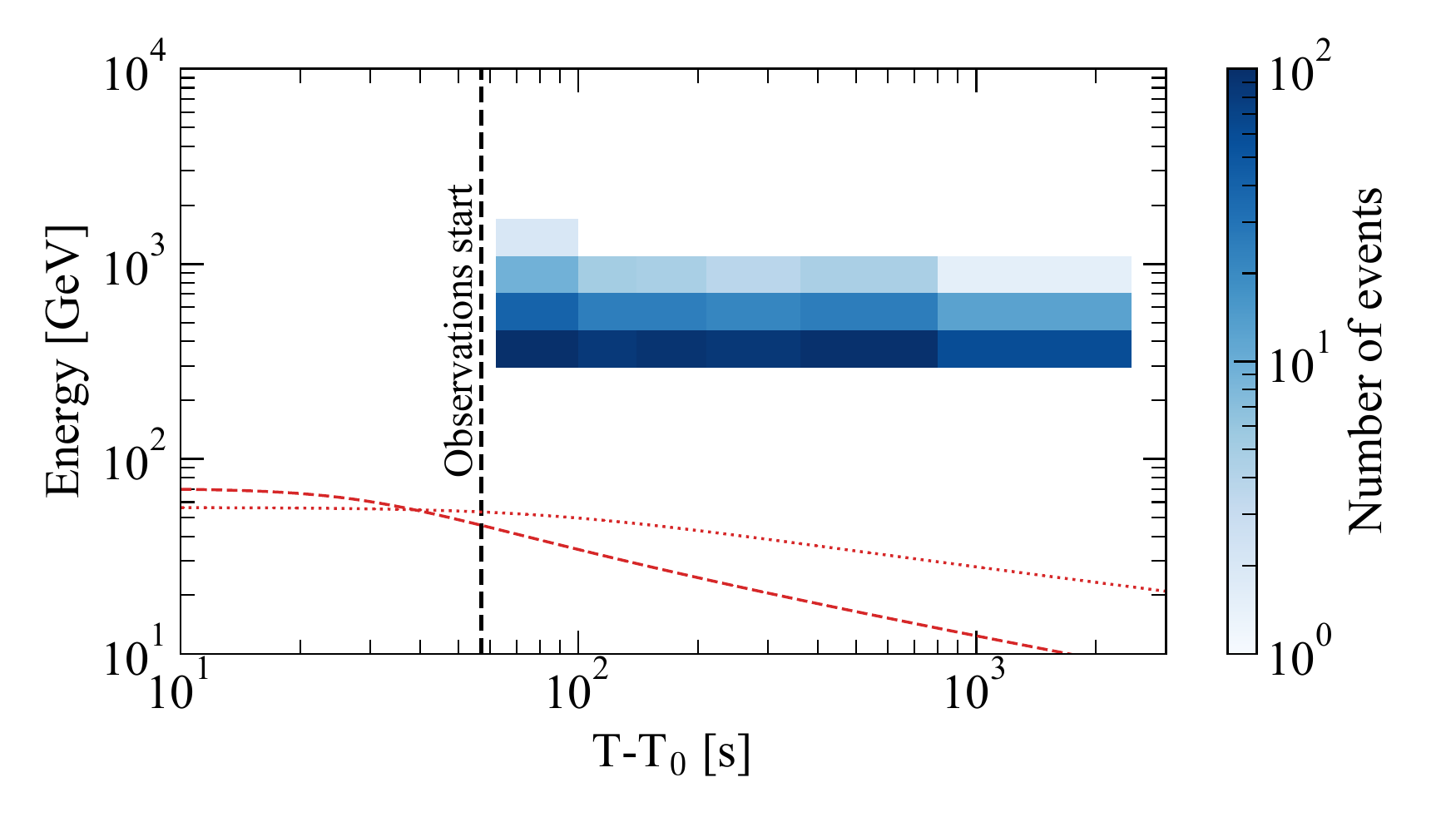}
\caption{
{\bf Distribution of TeV-band gamma rays in energy versus time for GRB 190114C.}
The number of events in each bin of energy and time are color-coded (Methods).
The vertical line indicates the beginning of data acquisition.
Curves show the expected maximum photon energy $\varepsilon_{\rm syn, max}$ of electron synchrotron radiation in the standard afterglow theory, for two extreme cases giving high values of $\varepsilon_{\rm syn, max}$. Dotted curve: isotropic-equivalent blast wave kinetic energy $E_{\rm k,aft}=3 \times 10^{55}\,{\rm erg}$ and homogeneous external medium with density $n=0.01\,{\rm cm^{-3}}$; dashed curve: $E_{\rm k,aft}=3 \times 10^{55}\,{\rm erg}$ and external medium describing a progenitor stellar wind with density profile $n(R)=A R^{-2}$ as function of radius $R$, where $A=3\times10^{33} \,{\rm cm^{-1}}$ (Methods).}
\label{fig:photonenergy-time}
\end{figure}

Much of the observed emission up to GeV energies for GRB 190114C is likely afterglow synchrotron emission from electrons, similar to many previous GRBs\cite{Ackermannetal2013,Kumar&Zhang2015}.
The TeV emission observed here is also plausibly associated with the afterglow.
However, it cannot be a simple spectral extension of the electron synchrotron emission.
The maximum energy of the emitting electrons is determined by the balance between their energy losses, dominated by synchrotron radiation, and their acceleration. The timescale of the latter should not be much shorter than the timescale of their gyration around the magnetic field at the external shock.
The energy of afterglow synchrotron photons is then limited to a maximum value, the so-called synchrotron burnoff limit\cite{Piran&Nakar2010,Ackermannetal2014} of $\varepsilon_{\rm syn, max} \sim 100 (\Gamma_b/1000)$ GeV, which depends only on the bulk Lorentz factor $\Gamma_b$. The latter is unlikely to significantly exceed $\Gamma_b \sim 1000$ (Methods).
Fig.~\ref{fig:photonenergy-time} compares the observed photon energies with expectations of $\varepsilon_{\rm syn, max}$ under different assumptions.
Although a few gamma rays with energy approaching $\varepsilon_{\rm syn, max}$ had been previously detected from a GRB by {\it Fermi}\cite{Ackermannetal2014}, the evidence for a separate spectral component was not conclusive, given the uncertainties in $\Gamma_b$, electron acceleration rate, and the spatial structure of the emitting region\cite{Kouveliotouetal2013}.
Here, even the lowest energy photons detected by MAGIC are significantly above $\varepsilon_{\rm syn, max}$ and extend beyond 1~TeV at 95\% confidence level (Methods).
Thus, this observation provides the first unequivocal evidence for a new emission component beyond synchrotron emission in the afterglow of a GRB.
Moreover, this component is energetically important, with power nearly comparable to that in the synchrotron component observed contemporaneously.

Comparing with previous MAGIC observations of GRBs,
the fact that GRB 190114C was the first to be clearly detected may be due to a favourable combination of its low redshift and suitable observing conditions rather than its intrinsic properties being exceptional (Methods),
although firm conclusions cannot yet be drawn with only one positive detection.
The capability of the telescopes to react fast and operate during moonlight conditions was crucial in achieving this detection.

The discovery of an energetically important emission component beyond electron synchrotron emission that may possibly be common in GRB afterglows offers important new insight into the physics of GRBs.
The similarity of the radiated power and temporal decay slopes in the TeV and X-ray bands suggests that this component is intimately related to the electron synchrotron emission.
Promising mechanisms for the TeV emission are ``leptonic'' processes in the afterglow
such as inverse Compton radiation, in which the electrons in the external shock Compton scatter ambient low-energy photons up to higher energies \cite{Meszarosetal1994,Zhang&Meszaros2001,Beniaminietal2015}.
On the other hand, ``hadronic'' processes induced by ultrahigh-energy protons in the external shock \cite{Vietri1997,Boettcher&Dermer1998,Zhang&Meszaros2001}
may also be viable if the acceleration of electrons and protons occur in a correlated manner.
However, such processes typically have low radiative efficiency, and are disfavoured as the origin of the luminous TeV emission observed in GRB 190114C for cases such as proton synchrotron emission (Methods).
Continuing efforts with existing and future gamma-ray telescopes will test these expectations and provide further insight into the physics of GRBs and related issues.

\newpage

\begin{methods}
\label{sec:methods}
\subsection{General properties of GRB\,190114C.}

GRB\,190114C was first identified by the {\it Swift}/BAT\cite{Gropp2019} and {\it Fermi}/GBM\cite{Hamburg2019} instruments on 14 January 2019, 20:57:03\,UT. Subsequently, it was also detected by several other space-based instruments, including {\it Fermi}/LAT, {\it INTEGRAL}/SPI-ACS, {\it AGILE}/MCAL, {\it Insight}/HXMT and Konus-Wind\cite{Ravasioetal2019}.
Its redshift was reported as $z = 0.4245 \pm 0.0005$ by the Nordic Optical Telescope\cite{Selsing2019} and confirmed by Gran Telescopio Canarias\cite{Castro-Tirado2019}.
The measured duration of $T_{90} \sim 116$\,s by {\it Fermi}/GBM and $T_{90} \sim 362$\,s by {\it Swift}/BAT\cite{Krimmetal2019} puts GRB\,190114C unambiguously in the long-duration subclass of GRBs\cite{Gehrels&Meszaros2012}.
The fluence and peak photon flux of the emission at $10-1000$\,keV during $T_{90}$ measured by GBM are $(3.990\pm0.008) \times 10^{-4}$\,erg\,cm$^{-2}$ and $(246.86 \pm 0.86)$\,ph\,cm$^{-2}$\,s$^{-1}$ \cite{Hamburg2019}, corresponding to $E_{\rm iso} \sim 3 \times 10^{53}$\,erg and $L_{\rm iso} \sim 1 \times 10^{53}$\,erg\,s$^{-1}$, respectively\cite{Ravasioetal2019}.
These values are consistent with the known correlations for GRBs between their spectral peak energy $\varepsilon_{\rm peak}$ and $E_{\rm iso}$\cite{Amatietal2002}, and between $\varepsilon_{\rm peak}$ and  $L_{\rm iso}$\cite{Yonetokuetal2004}.
The light curve of the keV-MeV emission exhibits two prominent emission episodes with irregular, multi-peaked structure, at $t \sim$ 0-5 s and $t \sim$ 15-25 s (Extended Data Fig.\ref{fig:early_magic_lc}).
The spectra for these episodes are typical of GRB prompt emission\cite{Ravasioetal2019}.
On the other hand, at $t \sim$ 15-25 s and $t \gtrsim$ 25 s, the temporal and spectral properties of the keV-MeV emission are consistent with an afterglow component, indicating a significant overlap in time of the prompt and afterglow phases.
Indeed, from a joint spectral and temporal analysis of the {\it Fermi}/GBM and /LAT data, the onset of the afterglow for GRB 190114C was estimated to occur at $t \sim 6$\,sec, much earlier than $T_{90}$\cite{Ravasioetal2019}.

The event is fairly energetic but not exceptionally so, with $E_{\rm iso}$ lying in the highest $\sim $30\% of its known distribution\cite{Navaetal2012}.
No neutrinos were detected by the IceCube Observatory in the energy range \SI{100}{\TeV} to \SI{10}{\peta\eV}, under non-optimal observing conditions\cite{Vandenbroucke2019}.

\subsection{MAGIC Telescopes and Automatic Alert System.}

The MAGIC telescopes comprise two 17-m diameter IACTs (MAGIC-I and MAGIC-II) operating in stereoscopic mode, located at the Roque de los Muchachos Observatory in La Palma, Canary Islands, Spain\cite{Aleksicetal2016a, Aleksicetal2016b}.
By imaging Cherenkov light from extended air shower events, the telescopes can detect gamma rays above an energy threshold of 30 GeV
depending on the observing mode and conditions, with a field of view of $\sim$10 square degrees.

Observing GRBs with IACTs such as MAGIC warrants a dedicated strategy.
As the probability of discovering GRBs by IACTs serendipitously in their relatively small field of view is low, they rely on external alerts provided by satellite instruments with larger fields of view to trigger follow-up observations.
Since their inception, the MAGIC telescopes were designed to perform fast follow-up observations of GRBs.
By virtue of their light-weight reinforced carbon fiber structure and high repositioning speed, they can respond quickly to GRB alerts received via the Gamma-ray Coordinates Network (GCN\footnote{\url{https://gcn.gsfc.nasa.gov}})\cite{Barthelmy2016}.
After various updates to the entire system over the years\cite{Aleksicetal2016a,Aleksicetal2016b}, the telescopes can currently slew to a target with a repositioning speed of 7 degrees per second.
To achieve the fastest possible response to GRB alerts, an Automatic Alert System (AAS) has been developed, which is a multi-threaded program that performs different tasks such as connecting to the GCN servers, receiving GCN Notices that contain the sky coordinates of the GRB, and sending commands to the Central Control (CC) software of the MAGIC telescopes.
This also includes a check of the visibility of the new target according to predefined criteria.
A priority list was set up for cases when several different types of alerts are received simultaneously.
Moreover, if there are multiple alerts for the same GRB, the AAS will select the one with the best localization.

If an alert is tagged as observable by the AAS, the telescopes will automatically repoint to the new sky position.
An automatic procedure, implemented in 2013, prepares the subsystems for data taking during the telescope slewing\cite{Carosietal2015,Bertietal2017}: data taken previously is saved, relevant trigger tables are loaded, appropriate
electronics thresholds are set and the mirror segments are suitably adjusted by the Automatic Mirror Control hardware.
While moving, the telescopes calibrate the imaging cameras. 
The Data Acquisition (DAQ) system continues taking data while it receives information about the target from the CC software.
The presence of a trigger limiter set to \SI{1}{\kilo\hertz} prevents high rates and the saturation of the DAQ system.
When the repositioning has finished, the target is tracked in wobble mode, which is the standard observing mode for MAGIC\cite{Fominetal1994}.
To date, the fastest GRB follow-up was achieved for GRB\,160821B, when the data taking started only 24 seconds after the GRB.

\subsection{MAGIC observations of GRB 190114C.}

On the night of 14 January 2019, at 20:57:25 UT ($T_0+\SI{22}{\second}$), {\it Swift}/BAT distributed an alert reporting the first estimated coordinates of GRB\,190114C (RA: +03h 38m 02s; Dec: -26d 56m 18s).
The AAS validated it as observable and triggered the automatic repointing procedure, and the telescopes began slewing in fast mode from the position before the alert. The MAGIC-I and MAGIC-II telescopes were on target and began tracking GRB\,190114C at 20:57:52.858 UT and 20:57:53.260 UT ($T_0+\SI{50}{\second}$), respectively, starting from zenith angle \ang{55.8} and azimuth angle \ang{175.1} in local coordinates.
After starting the slewing, the telescopes reached the target position in approximately 27 seconds, moving by 42.82 degrees in zenith and 177.5 degrees in azimuth.
At the end of the slewing, the cameras on the telescopes oscillated for a short time. Subsequently, we performed dedicated tests that reproduced the movement of the telescopes. We verified that the duration of the oscillations was less than 10 seconds after the start of tracking, and its amplitude was less than 0.6 arc-minutes when data taking began.
Data acquisition started at 20:58:00 ($T_0+\SI{57}{\second}$) and the DAQ system was operating stably from 20:58:05 ($T_0+\SI{62}{\second}$), as denoted in Extended Data Fig. 1.

Observations were performed in the presence of moonlight, implying a relatively high night sky background (NSB), approximately $\sim 6$ times the level for dark observations (moonless nights with good weather conditions)\cite{Ahnen2017}. 
Data taking for GRB\,190114C stopped on 15 January 2019, 01:22:15 UT, when the target reached zenith angle \ang{81.14} and azimuth angle \ang{232.6}. The total exposure time for GRB\,190114C was \SI{4.12}{\hour}. 

\subsection{MAGIC data analysis for GRB\,190114C.}

Data collected from GRB\,190114C were analysed using the standard MAGIC analysis software\cite{Aleksicetal2016b} and the analysis chain tuned for data taken under moonlight conditions\cite{Ahnen2017}.
No detailed information on the atmospheric transmission is available since the LIDAR facility\cite{Frucketal2014} was not operating during the night of the observation.
Therefore, the quality of the data was assessed by checking other auxiliary weather monitoring devices as well as the value and stability of the DAQ rates.

A dedicated set of Monte Carlo (MC) simulation gamma-ray data was produced for the analysis, matching the trigger settings (discriminator thresholds), the zenith-azimuth distribution, and the NSB level of GRB\,190114C observations.
The final data set comprises events starting from 20:58:05 UT. Due to the higher NSB, compared to standard analysis, a higher level of image cleaning was applied to both real and MC data, while a higher cut on the integrated charge of the event image, set to 80 photo-electrons, was used for evaluating photon fluxes\cite{Ahnen2017}. The significance of the gamma-ray signal was computed using the Li \& Ma method\cite{LiMa83}.

The spectra in \figurename\,\ref{fig:MAGIC_SED_timeinteg} were derived by assuming a simple power law function for the intrinsic spectrum,
\[
\frac{dF}{d\varepsilon}=f_{0}\times\left(\frac{\varepsilon}{\varepsilon_0}\right)^{-\alpha} ,
\]
with the forward-folding method to derive the best fit parameters and the Schmelling unfolding prescription for the spectral points\cite{Schmelling1994}, starting from the observed spectrum and correcting for EBL attenuation with the model of Dominguez et al.\cite{Dominguezetal11}.
The best fit values are $\alpha_{\rm int}=-2.22\,^{+0.23}_{-0.25}\,\textup{(stat)}\,^{+0.21}_{-0.26}\,\textup{(sys)}$ and $f_\textup{0,int}=[\,8.45\,^{+0.68}_{-0.65}\,\textup{(stat)}\,^{+4.42}_{-3.97}\,\textup{(sys)}\,]\cdot10^{-9}$ \si{\per\TeV\per\cm\squared\per\second} at \SI{0.46}{\TeV}.
Note that due to the soft spectrum of the source, the systematic errors reported here are larger than the ones given in Aleksic et al.\cite{Aleksicetal2016b}.

The absolute energy scale for MAGIC measurements is systematically affected by the imperfect knowledge of different aspects such as atmospheric transmission, mirror reflectance and properties of photomultipliers (PMTs) . A dedicated study\cite{Aleksicetal2016b} identified the light-scale matching of real and MC data as the most important contribution to the systematic errors on the absolute energy scale. A miscalibration of the MC energy scale can lead to mis-reconstruction of the spectrum that affect both the flux and its spectral slope, especially at the lowest energies. These studies demonstrated that the reconstructed spectra for MAGIC are affected by a systematic error due to the variation of the light scale by less than $\pm$15\%.
In the case of moonlight observations, additional systematic effects on the flux of the observed spectra arises from mismatches between MC and real data, in particular of the trigger discriminator thresholds (DTs) and of the higher noise in the PMTs. Dedicated studies for moonlight observations\cite{Ahnen2017} reveal that these errors affect only the overall spectral flux (and not the spectral index) and depend on the level of the NSB. The contribution to the systematic error coming from the moonlight observations is minor compared to that due to the light scale variations. Moreover, in the case of GRB\,190114C, the influence of moonlight conditions on the overall systematic errors is mitigated by the improved data-MC agreement achieved by simulating the recorded trigger DTs and NSB during the GRB\,190114C observation.
For the analysis of GRB\,190114C data, we reproduced the effect of the light scale variations on the spectra to derive the systematic errors on energy flux and the errors  on the photon index reported in Extended Data Table\,\ref{tab:energy_flux_lc}. The light scale modifications are applied to the spectra before their deconvolution with EBL attenuation, which finally affects the low and the high energy ends of the spectra in different ways. The fit to the obtained curves is performed in the same manner as the nominal case. Finally, the systematic errors are obtained from the difference of the parameter values computed for the nominal case and for the cases of light scale variations by $\pm$15\%.

An additional systematic effect originates from uncertainties in current EBL models. In order to quantify the corresponding systematic errors on the derived photon indices, the observed spectra were corrected by adopting several different EBL models\cite{Franceschinietal2008,Finkeetal2010,Gilmoreetal2012} for the redshift of this GRB. The results can be found in Extended Data Table\,\ref{tab:ebl_models_comparison}. The spectral indices inferred using different EBL models differ less than their statistical uncertainties (1 standard deviation). Taking as reference the EBL model by Dominguez et al.\cite{Dominguezetal11}, the spectral index for the time-integrated spectrum has an additional systematic error due to uncertainties in the EBL such that $\alpha_{\rm int}=-2.22\,^{+0.23}_{-0.25}\,\textup{(stat)}\,^{+0.21}_{-0.26}\,\textup{(sys)}\,^{+0.07}_{-0.17}\,\textup{(sys$_{\textup{EBL}}$)}$.
The observed spectrum in the $0.2-1.0$\,TeV energy range can be roughly described by a power-law  with photon index $\alpha_{\rm obs} = -5.43 \pm 0.22\,\textup{(stat)}$ and flux normalization $f_\textup{0,obs}=[\,4.09\pm0.34\,\textup{(stat)}\,]\cdot10^{-10}$ \si{\per\TeV\per\cm\squared\per\second} at $0.475$ TeV.

The upper limit for the first non-significant energy bin in the observed spectrum shown in Figure \ref{fig:MAGIC_SED_timeinteg} is calculated from a likelihood ratio test between two models. The first, baseline model only considers background events and spillover events from lower energy. The second model additionally assumes that the spectrum extends to higher energy as an unbroken power-law, with the flux normalization as a free parameter. Given the low event statistics in the higher energy bins, the validity of the upper limit was checked by performing 10,000 Monte Carlo simulations of the likelihood ratio test. The test statistic distribution derived from this toy simulation is then used to determine the upper limit on the flux at 95\% confidence level. The corresponding upper limit for the intrinsic spectrum is derived from that for the observed spectrum by correcting for EBL attenuation.

The time-dependent, EBL-corrected energy flux values shown in \figurename\,\ref{fig:full_MAGIC_lc} and reported in Extended Data Table\,\ref{tab:energy_flux_lc} were computed with an analytical procedure.
For each time bin, the value of the energy flux is computed as the integral between \num{0.3} and \SI{1}{\TeV} of the best-fit spectral power law function derived with the forward folding method. Accordingly, the errors are calculated analytically through standard procedures for error propagation, taking into account the covariance matrix. Moreover, the analytical results were checked against those computed with a toy Monte Carlo simulation, which gives comparable results.

The lower limits on the maximum event energy were computed by an iterative procedure where a power-law model is assumed for the intrinsic spectrum, and a different cut is applied to the maximum event energy for each iteration.
For each value of the energy cut, a forward-folding fit is performed and a $\chi^2$ value is obtained.
The final result is obtained by finding the value of the energy cut for which the $\chi^2$ variation corresponds to a given confidence level, set here to 95\%. 

The number of events in each time and energy bin shown in Figure\,\ref{fig:photonenergy-time} was computed using the forward-folding EBL-corrected spectrum, the instrument effective area and the effective time of the observation. For the highest energy bins, the corresponding numbers for the time interval between $T_0+\SI{62}{\second}$ and $T_0+\SI{1227}{\second}$ are listed in Extended Data Table\,\ref{tab:excess_counts}.

The number of observed excess events in bins of estimated energy are reported in Extended Data Table\,\ref{tab:observed_expected_photons}. Also listed are the expected number of photons in the same energy bins, obtained from the power-law model of the intrinsic spectrum by convolving it with the effect of EBL attenuation and the instrument response function for the zenith angles of this observation. Note that the counts in bins of estimated energy cannot be used to derive physical inferences.
Spectral information that is physically meaningful must be computed as a function of
the true energy of the events through an unfolding procedure using the energy migration matrix. Figure\,\ref{fig:MAGIC_SED_timeinteg} shows such unfolded spectra (both intrinsic and observed) as a function of the true event energies.

\subsection{Fermi/LAT data analysis for GRB~190114C.}
The publicly available Pass 8 (P8R3) LAT data for GRB~190114C was processed using the Conda fermitools v1.0.2 package, distributed by the Fermi collaboration\footnote{\url{https://fermi.gsfc.nasa.gov/ssc/data/analysis/software/}}. Events of the ``Transient'' class (P8R3\_TRANSIENT020\_V2) were selected within $10^\circ$ from the source position. We assumed a power law spectrum in the $0.1-10$~GeV energy range, also accounting for the diffuse galactic and extragalactic backgrounds, as described in the analysis manual\footnote{\url{https://fermi.gsfc.nasa.gov/ssc/data/analysis/scitools/}}. To compute the source fluxes, we first checked that the spectral index is consistent with $-2$ for the entire 62--180 seconds interval after $T_0$, and then repeated the fit, fixing the index to this value. The LAT energy flux shown in Fig.~\ref{fig:full_MAGIC_lc} was computed as the integral of the best-fit power law model within the corresponding energy range.

\subsection{XRT light curve.} 
The XRT lightcurve shown in Fig.\ref{fig:full_MAGIC_lc} was derived from the online analysis tool that is publicly available at the {\it Swift}-XRT repository\footnote{\url{http://www.swift.ac.uk/xrt_curves/}}. The spectral data collected in the Windowed Timing (WT) mode suffered from an instrumental effect, causing a non-physical excess of counts below $\sim0.8$\,keV\cite{XRT_GCN}. 
To remove this effect, we considered the best fit model of spectral data above 1\,keV and estimated a conversion factor from counts to deabsorbed flux equal to $10^{-10}$\,erg\,cm$^{-2}$\,ct$^{-1}$. This conversion factor was applied to the counts lightcurve to derive the energy flux light curve in the time interval 62-2000\,s.

\subsection{Synchrotron burnoff limit for the afterglow emission.}
GRB afterglows are triggered by external shocks that decelerate and dissipate their kinetic energy in the ambient medium, consequently producing a nonthermal distribution of electrons via mechanisms such as shock acceleration\cite{Kumar&Zhang2015}.
The maximum energy of electrons that can be attained in the reference frame comoving with the post-shock region can be estimated by equating the timescales of acceleration $\tau_{\rm acc}$ and energy loss $\tau_{\rm loss}$, the latter primarily due to synchrotron radiation\cite{Piran&Nakar2010}.
These are expected to scale with electron Lorentz factor $\gamma$ and magnetic field strength $B$ as $\tau_{\rm acc} \propto \gamma B^{-1}$ and $\tau_{\rm loss} \propto \gamma^{-1} B^{-2}$, so that the maximum electron Lorentz factor $\gamma_{\rm max}\propto B^{-1/2}$. 
Thus, the maximum energy of synchrotron emission $\varepsilon_{\rm syn,max}\propto B \gamma_{\rm max}^2$ is independent of $B$.
Its numerical value in the shock comoving frame is $\varepsilon^\prime_{\rm syn,max}\sim 50-100$\,MeV, determined only by fundamental constants and a factor of order one that characterizes uncertainties in the acceleration timescale.
The observed spectrum of afterglow synchrotron emission is then expected to display a cutoff below the energy $\varepsilon_{\rm syn,max} \sim 100 {\rm MeV} \times \Gamma_b(t) / (1+z)$, which depends only on the time-dependent bulk Lorentz factor $\Gamma_b(t)$ of the external shock.
To estimate $\varepsilon_{\rm syn,max}$ and its evolution, we employ $\Gamma_b(t)$ derived from solutions to the dynamical equations of the external shock \cite{Navaetal2013}.
The resulting curves for $\varepsilon_{\rm syn,max}$ are shown for cases of a medium with constant density $n=const$, and a medium with a radial density profile $n(R)= A\,R^{-2}$ (with $A=3\times 10^{35}\,A_\star\,$cm$^{-1}$), expected when a dense stellar wind is produced by the progenitor star (dotted and dashed lines in \figurename\,\ref{fig:photonenergy-time}, respectively). 
These curves have assumed small values for the density ($n=0.01$ and $A_\star=0.01$) and the efficiency of prompt emission ($\eta_\gamma=1\%$) that imply a large value for the isotropic-equivalent blastwave kinetic energy ($E_{\rm k,aft}=E_{\rm iso}(1-\eta_\gamma)/\eta_\gamma$), resulting in high values of $\varepsilon_{\rm syn,max}$.
Even with such extreme assumptions, the energy of photons detected by MAGIC are well above $\varepsilon_{\rm syn,max}$ (Fig.\ref{fig:photonenergy-time}).

\subsection{Constraints on proton synchrotron afterglow emission.}
Synchrotron emission by protons accelerated to ultrahigh-energies in the external shock has been proposed as a mechanism for GeV-TeV emission in GRB afterglows, potentially at energies above the burnoff limit for electron synchrotron emission\cite{Vietri1997,Boettcher&Dermer1998,Zhang&Meszaros2001,Totani1998,Razzaque2010}.
We discuss whether this process provides a viable explanation for the TeV emission observed here, following the formulation of Ref. 12. 
For the case of a uniform external medium with density $n=n_0 \ {\rm cm^{-3}}$, the maximum expected energy of proton synchrotron emission in the observer frame is
\begin{equation}
\varepsilon_{\rm psyn,max} = 7.6\, {\rm GeV}\, \eta^{-2}\, \epsilon_B^{3/2}\, (n_0\, E_{\rm k,53})^{3/4}\, t_s^{-1/4}\, (1+z)^{-3/4} ,
\end{equation}
where $E_{\rm k,aft}=10^{53}\, E_{\rm k,53}\, {\rm erg}$, $t_s$ is the observer time after the burst in seconds, $\epsilon_B$ is the fraction of energy in magnetic fields relative to that dissipated behind the shock, and $\eta$ is a factor of order one that characterizes the acceleration timescale.
Even when assuming optimistic values of $\epsilon_B=0.5$ and $\eta=1$, realising $\varepsilon_{\rm psyn,max} \gtrsim 1$ TeV at $t \sim 100$ s for a GRB at $z=0.42$ requires $n_0\,E_{\rm k,53} \gtrsim 10^4$, a very high value for the product of the blastwave energy and external medium density.

Even more severe is the requirement to reproduce the observed TeV flux and spectrum.
Assuming a power-law energy distribution with index $-p$ for the accelerated protons, their synchrotron emission is expected to have a single power-law spectrum with photon index $\alpha_{\rm int} =-(p+1)/2$, extending from a minimum energy
\begin{equation}
\varepsilon_m = 3.7 \times 10^{-3}\, {\rm eV}\, \xi_p^{-2}\, \epsilon_p^2\, \epsilon_B^{1/2}\, E_{\rm k,53}^{1/2}\, t_s^{-3/2}\, (1+z)^{1/2}
\end{equation}
with differential energy flux
\begin{equation}
f(\varepsilon=\varepsilon_m) = 1.3 \times 10^{-28}\, {\rm erg\, cm^{-2}\, s^{-1}\, Hz^{-1}} \xi_p\, \epsilon_B^{1/2}\, n_0^{1/2}\, E_{\rm k,53}\, D_{28}^{-2}\, (1+z)
\end{equation}
up to $\varepsilon = \varepsilon_{\rm psyn,max}$,
where $\xi_p$ is the fraction in number of the protons swept up by the shock that are accelerated, $\epsilon_p$ is the fraction of energy in accelerated protons relative to that dissipated behind the shock, and $D= 10^{28}\, D_{28}\, {\rm cm}$ is the luminosity distance of the GRB.
The observed intrinsic spectral index $\alpha_{\rm int} \sim -2$ at $t \sim 100$ s implies $p \sim 3$.
If $p=3$ and the spectrum extends to $\varepsilon=$1 TeV without a cutoff, the energy flux at 1 TeV is
\begin{equation}
F(\varepsilon=1\, {\rm TeV}) = 1.1 \times 10^{-16}\, {\rm erg\, cm^{-2}\, s^{-1}}\, \epsilon_p^2\, \xi_p^{-1}\, \epsilon_B\, n_0^{1/2}\, E_{\rm k,53}^{3/2}\, D_{28}^{-2}\, t_s^{-3/2}\, (1+z)^{3/2} .
\end{equation}
With optimistic assumptions of $\epsilon_B=0.5$, $\eta=1$, $\epsilon_p=0.5$ and $\xi_p=0.1$, accounting for the observed 0.3-1 TeV flux at $t \sim 100$ s of $F \sim 4 \times 10^{-8}\, {\rm erg\, cm^{-2}\, s^{-1}}$ necessitates $n_0^{1/2}\, E_{\rm k,53}^{3/2} \gtrsim 10^{11}$.
Even in the extreme case of a GRB occurring at the center of a dense molecular cloud with $n=10^6\, {\rm cm^{-3}}$, the blastwave energy must be $E_{\rm k,aft} > 2 \times 10^{59}\, {\rm erg}$, far exceeding the energy available for any plausible GRB progenitor\cite{Kumar&Zhang2015}.
This conclusion is qualitatively valid regardless of how the electron synchrotron emission is modelled, or if the external medium has a density profile characteristic of a progenitor stellar wind.
Although proton synchrotron emission may possibly explain the GeV emission observed in some GRBs\cite{Razzaque2010}, due to its low radiative efficiency, it is strongly disfavoured as the origin of the luminous TeV emission observed in GRB 190114C.
A more plausible mechanism may be inverse Compton emission by accelerated electrons\cite{Meszarosetal1994,Zhang&Meszaros2001,Beniaminietal2015,Galli&Piro2008}.

\subsection{Past TeV-band observations of GRBs with MAGIC and other facilities.}
The search for TeV gamma rays from GRBs had been pursued over many years employing a variety of experimental techniques, but no clear detections had been previously achieved
\cite{Connaughtonetal1999,Atkinsetal2000,Atkinsetal2004,Abdoetal2007,Horanetal2007,Aharonianetal2009a,Aharonianetal2009b,Acciarietal2011,Abramowskietal2014,Alfaroetal2017,Hoischenetal2017,Abeysekaraetal2018}.

Designed with GRB follow-up observations as a primary goal, MAGIC has been responding to GRB alerts since 15th July 2004.
For the first 5 years, MAGIC operated as a single telescope (MAGIC-I), reacting mainly to alerts from \textit{Swift}.
After the second telescope (MAGIC-II) was added in 2009, GRB observations have been carried out in stereoscopic mode.
Excluding cases when proper data could not be taken due to hardware problems or weather conditions, 105 GRBs were observed from July 2004 to February 2019.
Of these, 40 have determined redshifts, among which 8 and 3 have redshifts lower than 1 and 0.5, respectively.
Observations started less than 30 minutes after the burst for 66 events (of which 33 lack redshifts), and less than 60 seconds for 14 events.
The small number of the latter is mainly due to bad weather conditions or observational criteria that were not fulfilled at the time of the alert. 

Despite 15 years of dedicated efforts, no unambiguous evidence for gamma-ray signals from GRBs had been seen by MAGIC before GRB 190114C.
The flux upper limits for GRBs observed in 2005-2006 were found to be consistent with simple power-law extrapolations of their low-energy spectra when EBL attenuation was taken into account\cite{Albertetal2007}.
More detailed studies were presented for GRB\,080430\cite{Aleksicetal2010} and GRB\,090102\cite{Aleksicetal2014} that were simultaneously observed with MAGIC and other instruments in different energy bands.
Since 2013, GRB observations have been performed with the new automatic procedure described above\cite{Carosietal2015,Bertietal2017}.
In addition, for some bright GRBs detected by {Fermi}/LAT, late-time observations have been conducted up to one day after the burst to search for potential signals extended in time.

The case of GRB\,190114C can be compared with other GRBs followed up by MAGIC under similar conditions.
Aside from the intrinsic spectrum, the main factors affecting the detectability of a GRB by IACTs are the redshift $z$ (stronger EBL attenuation for higher $z$), the zenith distance (higher energy threshold for higher zenith distance), outside light conditions and the delay time $T_{\rm delay}$ between the GRB and the beginning of the observations. If we select GRBs with $z<1$ and $T_{\rm delay} <\SI{1}{\hour}$, only four events remain, as listed in Extended Data Table\,\ref{tab:GRBs}.
Except for GRB\,190114C, these are all short GRBs, which is not surprising as they are known to be distributed at redshifts appreciably lower than long GRBs \cite{Ghirlandaetal2016}.
A few other long GRBs with $z<1$ were actually followed up by MAGIC with $T_{\rm delay} <\SI{1}{\hour}$, but the observations were not successful due to technical problems or adverse observing conditions.
There is also a fair fraction of events without measured redshifts.
Assuming that they follow the known $z$ distribution of long GRBs, $\sim 20$\% of the events are expected at $z < 1$\cite{Perleyetal2016}.
Since 30 long GRBs without redshifts were observed by MAGIC with $T_{\rm delay} <\SI{1}{\hour}$, the total number of events with observing conditions and $z$ similar to GRB\,190114C during the whole MAGIC GRB campaign is likely to be only a few.

A similar analysis for past GRBs observed by other Cherenkov telescopes is not possible, since not all the relevant ancillary information is available. However, summaries of the past efforts have been reported. Of the 150 GRBs followed up by VERITAS till February 2018\cite{Abeysekaraetal2018}, 50 had observations starting within 180\,sec from the satellite trigger time. H.E.S.S. also conducted several tens of GRB follow-up observations till 2017\cite{Aharonianetal2009b,Lennarzetal2013}. 64 GRBs were observed by HAWC\cite{Alfaroetal2017} till February 2017. Milagrito and Milagro observed 54 GRBs from February 1997 to May 1998\cite{Smithetal2000} and more than 130 GRBs from January 2000 to March 2008, respectively\cite{Aune2009,SazParkinson2009}. None of these considerable observational efforts provided any convincing detection, although some hints at low significance have been found. A case of particular interest was the Milagrito result for GRB\,970417A\cite{Atkinsetal2000}, although its statistical significance was not high enough to fully rule out a background event.
\end{methods}

\begin{addendum}
 \item [Acknowledgements]
 We are grateful to the referees for their constructive remarks that helped us improve the format and content of this manuscript.
 We dedicate this paper to the memory of Eckart Lorenz. With his innovative spirit, infinite enthusiasm and vast knowledge of experimental methods, techniques and materials, he played a key role in optimising the design of MAGIC, specifically for observations of GRBs. 
  We would like to thank the Instituto de Astrof\'{\i}sica de Canarias for the excellent working conditions at the Observatorio del Roque de los Muchachos in La Palma. The financial support of the German BMBF and MPG, the Italian INFN and INAF, the Swiss National Fund SNF, the ERDF under the Spanish MINECO (FPA2017-87859-P, FPA2017-85668-P, FPA2017-82729-C6-2-R, FPA2017-82729-C6-6-R, FPA2017-82729-C6-5-R, AYA2015-71042-P, AYA2016-76012-C3-1-P, ESP2017-87055-C2-2-P, FPA2017‐90566‐REDC), the Indian Department of Atomic Energy, the Japanese JSPS and MEXT, the Bulgarian Ministry of Education and Science, National RI Roadmap Project DO1-153/28.08.2018 and the Academy of Finland grant nr. 320045 is gratefully acknowledged. This work was also supported by the Spanish Centro de Excelencia ``Severo Ochoa'' SEV-2016-0588 and SEV-2015-0548, and Unidad de Excelencia ``Mar\'{\i}a de Maeztu'' MDM-2014-0369, by the Croatian Science Foundation (HrZZ) Project IP-2016-06-9782 and the University of Rijeka Project 13.12.1.3.02, by the DFG Collaborative Research Centers SFB823/C4 and SFB876/C3, the Polish National Research Centre grant UMO-2016/22/M/ST9/00382 and by the Brazilian MCTIC, CNPq and FAPERJ.
S. Inoue is supported by JSPS KAKENHI Grant Number JP17K05460, MEXT, Japan, and the RIKEN iTHEMS Program.
L. Nava acknowledges funding from the European Union's Horizon 2020 Research and Innovation programme under the Marie Sk\l odowska-Curie grant agreement n.\,664931.
K. Noda is supported by JSPS KAKENHI Grant Number JP19K21043, MEXT, Japan.
A. Berti acknowledges support from the Physics Department of the University
of Torino (through the \textit{Department of Excellence} funding) and from the Torino division of the Italian INFN.
E. Moretti acknowledges funding from the European Union’s Horizon 2020 research and innovation programme under Marie Sk\l odowska-Curie grant agreement No 665919.

\item[Competing Interests] The authors declare that they have no
competing financial interests.
\item[Author Contributions] The MAGIC telescope system was designed and constructed by the MAGIC Collaboration. Operation, data processing, calibration, Monte Carlo simulations of the detector,  and of theoretical models, and data analyses were performed by the members of the MAGIC Collaboration, who also discussed and approved the scientific results. All MAGIC collaborators contributed to the editing and comments to the final version of the manuscript. Susumu Inoue and Lara Nava coordinated the interpretation of the data, and together with Stefano Covino, wrote the corresponding sections, and contributed to the structuring and editing of the rest of the paper.  Koji Noda and Alessio Berti coordinated the analysis of the MAGIC data. Together with Elena Moretti they contributed in the analysis and the writing of the relevant sections. Ievgen Vovk performed Fermi/LAT analysis, and together with Davide Miceli contributed to the calculation of limits, excesses and to the curves in Fig. 3. Razmik Mirzoyan contributed in coordinating, structuring and editing this paper.
\item[Author Information] Reprints and Permissions information is available at www.nature.com/reprints. The authors declare no competing financial interest. Readers are welcome to comment on the online version of the paper. Correspondence and requests for materials should be addressed to Razmik Mirzoyan (Razmik.Mirzoyan@mpp.mpg.de).
\item[Data Availability Statement] Raw data were generated at the MAGIC telescopes large-scale facility. Derived data supporting the findings of this study are available from the corresponding authors upon request.
\item[Code Availability Statement] Proprietary data reconstruction codes were generated at the MAGIC telescopes large-scale facility. Information supporting the findings of this study are available from the corresponding authors upon request.
\end{addendum}

\renewcommand{\figurename}{Extended Data Figure}
\setcounter{figure}{0}
\renewcommand{\tablename}{Extended Data Table}

\begin{figure}
\centering
\includegraphics[width=0.9\textwidth]{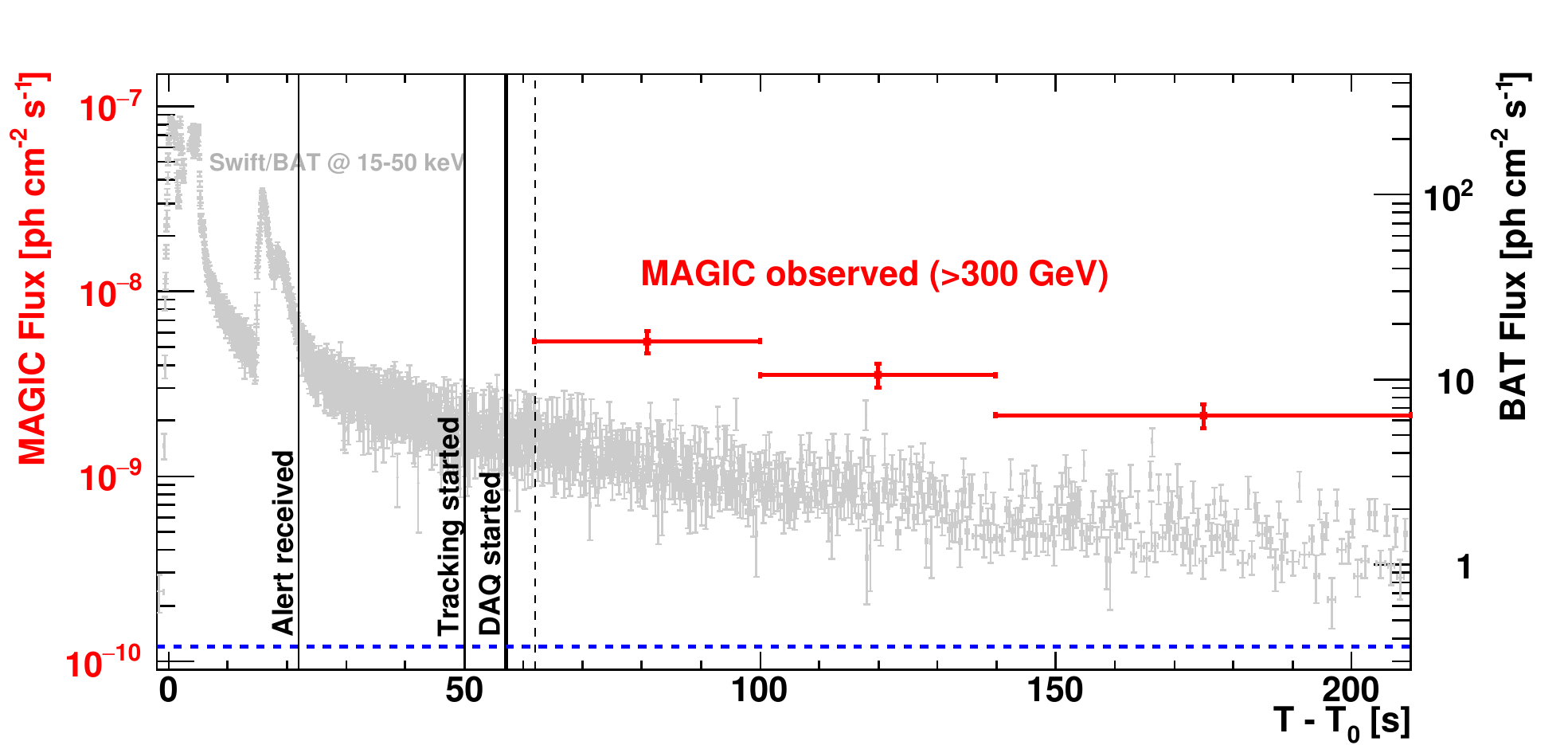}
\caption[Light curves in the TeV and keV bands for GRB 190114C.]{
{\bf Light curves in the TeV and keV bands for GRB 190114C.}
Light curve above \SI{0.3}{\TeV} in photon flux measured by MAGIC (red, from $T_0+\SI{62}{\second}$ to $T_0+\SI{210}{\second}$), compared with that between \SI{15}{\keV} and \SI{50}{\keV} measured by {\it Swift}/BAT\cite{Evansetal2010} (grey, from $T_0$ to $T_0+\SI{210}{\second}$) and the photon flux above \SI{0.3}{\TeV} of the Crab Nebula (blue dashed line). The errors on the MAGIC photon fluxes correspond to 1 standard deviation. Vertical lines indicate the times for MAGIC when the alert was received ($T_0+\SI{22}{\second}$), when the tracking of the GRB by the telescopes started ($T_0+\SI{50}{\second}$), when the data acquisition started ($T_0+\SI{57}{\second}$), and when the data acquisition system became stable ($T_0+\SI{62}{\second}$, dotted line).}
\label{fig:early_magic_lc}
\end{figure}

\begin{figure}
\centering
\includegraphics[width=0.8\textwidth]{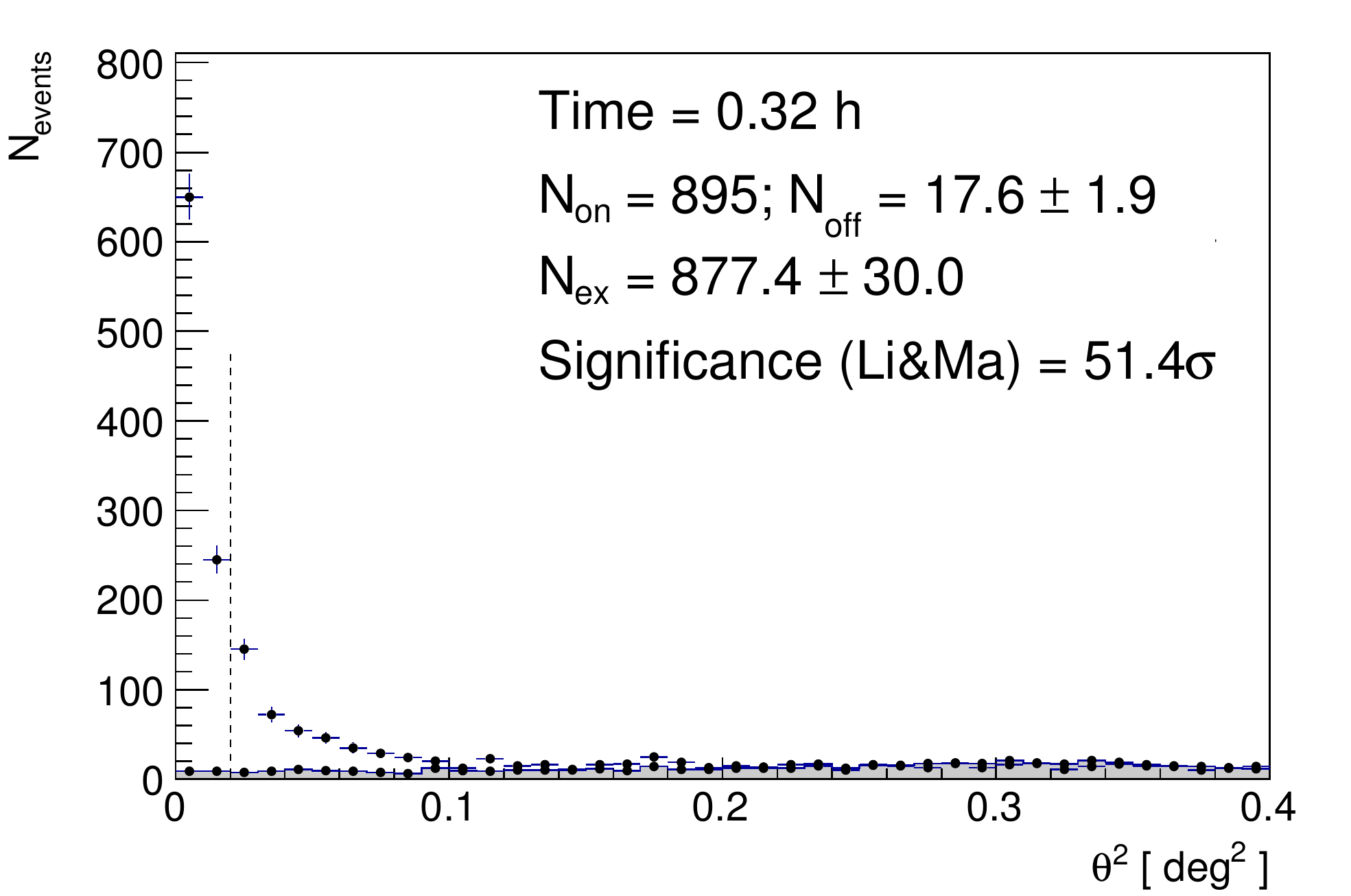}
\caption{
{\bf Significance of the gamma-ray signal between $T_0$+62 seconds and $T_0$+1227 seconds for GRB\,190114C.}
Distribution of the squared angular distance $\theta^2$ for the MAGIC data (points) and background events (grey shaded area). $\theta^2$ is defined as the squared angular distance between the nominal position of the source and the reconstructed arrival direction of the events. The dashed vertical line represents the value of the cut on $\theta^2$. This defines the signal region, where the number of events coming from the source ($N_\textup{ON}$) and from the background ($N_\textup{OFF}$) are computed. The errors for ON events are derived from the Poissonian statistics.
}
\label{fig:grb190114c_theta2_first_20min}
\end{figure}

\begin{table}
\centering
\begin{tabular}[th!]{ccc}
\toprule
Time bin & Energy flux & Spectral index  \\\relax
[\,seconds after $T_0$\,] & [\,\si{erg\,\cm^{-2}\,\second^{-1}}\,] & \\
\midrule
$62$ - $100$    & $[\,5.64 \pm 0.90\,\textup{(stat)}\,^{+3.24}_{-3.22}\,\textup{(sys)}\,]\cdot 10^{-8}$    & $-1.86\,^{+0.36}_{-0.40}\,\textup{(stat)}\,^{+0.12}_{-0.21}\,\textup{(sys)}$ \\
$100$ - $140$   & $[\,3.31 \pm 0.67\,\textup{(stat)}\,^{+2.71}_{-1.84}\,\textup{(sys)}\,]\cdot 10^{-8}$    & $-2.15\,^{+0.43}_{-0.48}\,\textup{(stat)}\,^{+0.25}_{-0.32}\,\textup{(sys)}$ \\
$140$ - $210$   & $[\,1.89 \pm 0.36\,\textup{(stat)}\,^{+1.72}_{-0.94}\,\textup{(sys)}\,]\cdot 10^{-8}$    & $-2.31\,^{+0.47}_{-0.54}\,\textup{(stat)}\,^{+0.15}_{-0.22}\,\textup{(sys)}$ \\
$210$ - $361.5$ & $[\,7.54 \pm 1.60\,\textup{(stat)}\,^{+6.46}_{-4.41}\,\textup{(sys)}\,]\cdot 10^{-9}$    & $-2.53\,^{+0.53}_{-0.62}\,\textup{(stat)}\,^{+0.22}_{-0.24}\,\textup{(sys)}$ \\
$361.5$ - $800$ & $[\,3.10 \pm 0.70\,\textup{(stat)}\,^{+1.20}_{-2.36}\,\textup{(sys)}\,]\cdot 10^{-9}$    & $-2.41\,^{+0.51}_{-0.65}\,\textup{(stat)}\,^{+0.27}_{-0.34}\,\textup{(sys)}$ \\
$800$ - $2454$  & $[\,4.54 \pm 2.04\,\textup{(stat)}\,^{+7.66}_{-1.96}\,\textup{(sys)}\,]\cdot 10^{-10}$   & $-3.10\,^{+0.87}_{-1.25}\,\textup{(stat)}\,^{+0.75}_{-0.24}\,\textup{(sys)}$ \\
$62-2454$ (time integrated) & - &$-2.22\,^{+0.23}_{-0.25}\,\textup{(stat)}\,^{+0.21}_{-0.26}\,\textup{(sys)}$ \\
\bottomrule
\end{tabular}
\caption{
{\bf Energy flux between \num{0.3} and \SI{1}{\TeV} in selected time bins for GRB 190114C.} 
Values are listed corresponding to the light curve in Figure\,\ref{fig:full_MAGIC_lc}. For each time bin, columns represent a) start time and end time of the bin; b) EBL-corrected energy flux in the \num{0.3}-\SI{1}{\TeV} range; c) best-fit spectral photon indices. The last row reports the value of the intrinsic spectral index for the time-integrated spectrum (Figure\,\ref{fig:MAGIC_SED_timeinteg}).  The reported statistical errors correspond to 1 standard deviation, while systematic errors are derived from the variation of the light scale by $\pm$15\% (see Methods).}
\label{tab:energy_flux_lc}
\end{table}

\begin{table}
    \centering
    \begin{tabular}[th!]{cccc}
        \toprule
        $E_\textup{min}$~[TeV] & $E_\textup{max}$~[TeV] & Model counts in $[E_\textup{min};E_\textup{max}]$ & Significance above $E_\textup{min}$ \\
        \midrule
        $0.71$ & $1.10$ &   $25.4$ & $5.8$ \\
        $1.10$ & $1.70$ &   $4.1$  & $2.5$ \\
        $1.70$ & $2.64$ &   $0.9$  & $1.5$ \\
        $2.64$ & $4.09$ &   $0.1$  & $0.1$ \\
        \bottomrule
    \end{tabular}
    \caption{{\bf Highest-energy counts from GRB 190114C}. The counts are estimated from the MAGIC data using the power law spectral model for the time interval between $T_0$+62 seconds and $T_0$+1227 seconds.}
    \label{tab:excess_counts}
\end{table}

\begin{table}
\centering
\begin{tabular}[ht!]{cccc}
\toprule
$E_\textup{est,min}$ [TeV] & $E_\textup{est,max}$ [TeV] & Observed photons  & Expected photons  \\
\midrule
$0.19$ & $0.29$ & $155 \pm 13$ & $219 \pm 73$ \\
$0.29$ & $0.46$ & $598 \pm 26$ & $564 \pm 53$ \\
$0.46$ & $0.71$ & $154 \pm 13$ & $180 \pm 16$ \\
$0.71$ & $1.10$ & $32 \pm 6$  & $28 \pm 3$  \\
$1.10$ & $1.70$ & $6.0 \pm 2.9$ & $5.6 \pm 0.4$   \\
$1.70$ & $2.64$ & $2.3 \pm 1.8$ & $1.2 \pm 0.1$   \\
\bottomrule
\end{tabular}
\caption{{\bf Observed and expected number of events in bins of estimated energy for GRB\,190114C}. The number of expected events is calculated starting from the intrinsic spectrum power-law model, convolving it with the effect of EBL attenuation and the instrument response function of the telescope for these large zenith angles. The energy binning in estimated energy matches the one in true energy (after unfolding) shown in Figure\,\ref{fig:MAGIC_SED_timeinteg} and Extended Data Table\,\ref{tab:excess_counts}. The large uncertainty in the number of expected events in the lowest energy bin is dominated by the uncertainty in the very low effective area of the telescopes close to the energy threshold of this analysis. The numbers reported in this table cannot be directly used for any physical inference. The measured spectrum needs to first be unfolded using the energy migration matrix\cite{Aleksicetal2016b}.}
\label{tab:observed_expected_photons}
\end{table}

\begin{table}
\centering
\begin{tabular}[th!]{ccccc}
\toprule
Time bin & D11 & F08 & FI10 & G12 \\\relax
[\,seconds after $T_0$\,] & & & & \\
\midrule
$62-100$    & $-1.86^{+0.36}_{-0.40}$ & $-2.04^{+0.36}_{-0.40}$ & $-1.81^{+0.36}_{-0.40}$ & $-1.95^{+0.36}_{-0.39}$ \\
$100-140$   & $-2.15^{+0.43}_{-0.48}$ & $-2.32^{+0.43}_{-0.48}$ & $-2.09^{+0.43}_{-0.48}$ & $-2.23^{+0.42}_{-0.48}$ \\
$140-210$   & $-2.31^{+0.47}_{-0.54}$ & $-2.48^{+0.47}_{-0.54}$ & $-2.25^{+0.47}_{-0.54}$ & $-2.39^{+0.47}_{-0.53}$ \\
$210-361.5$ & $-2.53^{+0.53}_{-0.62}$ & $-2.69^{+0.52}_{-0.61}$ & $-2.46^{+0.52}_{-0.61}$ & $-2.60^{+0.52}_{-0.61}$ \\
$361.5-800$ & $-2.41^{+0.51}_{-0.65}$ & $-2.58^{+0.51}_{-0.64}$ & $-2.34^{+0.51}_{-0.64}$ & $-2.49^{+0.51}_{-0.64}$ \\
$800-2454$  & $-3.10^{+0.87}_{-1.25}$ & $-3.20^{+0.83}_{-1.20}$ & $-2.96^{+0.83}_{-1.20}$ & $-3.08^{+0.82}_{-1.19}$ \\
$62-2454$ (time integrated) & $-2.22^{+0.23}_{-0.25}$ & $-2.39^{+0.23}_{-0.25}$ & $-2.15^{+0.23}_{-0.25}$ & $-2.29^{+0.23}_{-0.24}$ \\
\bottomrule
\end{tabular}
\caption[Spectral indices for different EBL models.]{
{\bf Spectral indices for different EBL models.} The abbreviations refer to the different EBL model adopted in each case. D11: Dominguez et al. 2011\cite{Dominguezetal11} (reported also in Extended Data Table\,\ref{tab:energy_flux_lc}); F08: Franceschini et al. 2008\cite{Franceschinietal2008}; FI10: Finke et al. 2010\cite{Finkeetal2010}; G12: Gilmore et al. 2012\cite{Gilmoreetal2012}. The errors reported for the indices correspond to 1 standard deviation.}
\label{tab:ebl_models_comparison}
\end{table}

\begin{table}
\centering
\begin{tabular}[th!]{cccc}
\toprule
Event & redshift & $T_\textup{delay}$ (s) & Zenith angle (deg) \\
\midrule
GRB\,061217     &   $0.83$   &  $786.0$    &   $59.9$ \\
GRB\,100816A    &   $0.80$   &  $1439.0$   &   $26.0$ \\
GRB\,160821B    &   $0.16$   &  $24.0$     &   $34.0$ \\
GRB\,190114C    &   $0.42$   &  $58.0$     &   $55.8$ \\
\bottomrule
\end{tabular}
\caption{
{\bf List of GRBs observed under adequate technical and weather conditions by MAGIC with $z<1$ and $T_{\rm delay} <1$ h}. The zenith angle at the beginning of the observations is reported in the last column. All except GRB\,061217 were observed in stereoscopic mode. GRB\,061217, GRB\,100816A and GRB\,160821B are short GRBs, while GRB\,190114C is a long GRB. Observations for a few other long GRBs with the same criteria were also conducted but are not listed here, as they were affected by technical problems or adverse observing conditions.}
\label{tab:GRBs}
\end{table}


\begin{thebibliography}{10}
\expandafter\ifx\csname url\endcsname\relax
  \def\url#1{\texttt{#1}}\fi
\expandafter\ifx\csname urlprefix\endcsname\relax\def\urlprefix{URL }\fi
\providecommand{\bibinfo}[2]{#2}
\providecommand{\eprint}[2][]{\url{#2}}

\bibitem{Gehrels&Meszaros2012}
\bibinfo{author}{{Gehrels}, N.} \& \bibinfo{author}{{M{\'e}sz{\'a}ros}, P.}
\newblock \bibinfo{title}{{Gamma-Ray Bursts}}.
\newblock \emph{\bibinfo{journal}{Science}} \textbf{\bibinfo{volume}{337}},
  \bibinfo{pages}{932} (\bibinfo{year}{2012}).

\bibitem{Kumar&Zhang2015}
\bibinfo{author}{{Kumar}, P.} \& \bibinfo{author}{{Zhang}, B.}
\newblock \bibinfo{title}{{The physics of gamma-ray bursts \& relativistic
  jets}}.
\newblock \emph{\bibinfo{journal}{\physrep}} \textbf{\bibinfo{volume}{561}},
  \bibinfo{pages}{1--109} (\bibinfo{year}{2015}).

\bibitem{Meszaros2002}
\bibinfo{author}{{M{\'e}sz{\'a}ros}, P.}
\newblock \bibinfo{title}{{Theories of Gamma-Ray Bursts}}.
\newblock \emph{\bibinfo{journal}{\araa}} \textbf{\bibinfo{volume}{40}},
  \bibinfo{pages}{137--169} (\bibinfo{year}{2002}).

\bibitem{Piran2004}
\bibinfo{author}{{Piran}, T.}
\newblock \bibinfo{title}{{The physics of gamma-ray bursts}}.
\newblock \emph{\bibinfo{journal}{Reviews of Modern Physics}}
  \textbf{\bibinfo{volume}{76}}, \bibinfo{pages}{1143--1210}
  (\bibinfo{year}{2004}).

\bibitem{Meszarosetal2004}
\bibinfo{author}{{M{\'e}sz{\'a}ros}, P.}, \bibinfo{author}{{Razzaque}, S.} \&
  \bibinfo{author}{{Zhang}, B.}
\newblock \bibinfo{title}{{GeV-TeV emission from {$\gamma$}-ray bursts}}.
\newblock \emph{\bibinfo{journal}{\nar}} \textbf{\bibinfo{volume}{48}},
  \bibinfo{pages}{445--451} (\bibinfo{year}{2004}).

\bibitem{Fan&Piran2008}
\bibinfo{author}{{Fan}, Y.-Z.} \& \bibinfo{author}{{Piran}, T.}
\newblock \bibinfo{title}{{High-energy {$\gamma$}-ray emission from gamma-ray
  bursts --- before GLAST}}.
\newblock \emph{\bibinfo{journal}{Frontiers of Physics in China}}
  \textbf{\bibinfo{volume}{3}}, \bibinfo{pages}{306--330}
  (\bibinfo{year}{2008}).

\bibitem{Inoueetal2013}
\bibinfo{author}{{Inoue}, S.} \emph{et~al.}
\newblock \bibinfo{title}{{Gamma-ray burst science in the era of the Cherenkov
  Telescope Array}}.
\newblock \emph{\bibinfo{journal}{\app}} \textbf{\bibinfo{volume}{43}},
  \bibinfo{pages}{252--275} (\bibinfo{year}{2013}).

\bibitem{Nava2018}
\bibinfo{author}{{Nava}, L.}
\newblock \bibinfo{title}{{High-energy emission from gamma-ray bursts}}.
\newblock \emph{\bibinfo{journal}{International Journal of Modern Physics D}}
  \textbf{\bibinfo{volume}{27}}, \bibinfo{pages}{1842003}
  (\bibinfo{year}{2018}).

\bibitem{Meszarosetal1994}
\bibinfo{author}{{Meszaros}, P.}, \bibinfo{author}{{Rees}, M.~J.} \&
  \bibinfo{author}{{Papathanassiou}, H.}
\newblock \bibinfo{title}{{Spectral properties of blast-wave models of
  gamma-ray burst sources}}.
\newblock \emph{\bibinfo{journal}{\apj}} \textbf{\bibinfo{volume}{432}},
  \bibinfo{pages}{181--193} (\bibinfo{year}{1994}).

\bibitem{Zhang&Meszaros2001}
\bibinfo{author}{{Zhang}, B.} \& \bibinfo{author}{{M{\'e}sz{\'a}ros}, P.}
\newblock \bibinfo{title}{{High-Energy Spectral Components in Gamma-Ray Burst
  Afterglows}}.
\newblock \emph{\bibinfo{journal}{\apj}} \textbf{\bibinfo{volume}{559}},
  \bibinfo{pages}{110--122} (\bibinfo{year}{2001}).

\bibitem{Beniaminietal2015}
\bibinfo{author}{{Beniamini}, P.}, \bibinfo{author}{{Nava}, L.},
  \bibinfo{author}{{Duran}, R.~B.} \& \bibinfo{author}{{Piran}, T.}
\newblock \bibinfo{title}{{Energies of GRB blast waves and prompt efficiencies
  as implied by modelling of X-ray and GeV afterglows}}.
\newblock \emph{\bibinfo{journal}{\mnras}} \textbf{\bibinfo{volume}{454}},
  \bibinfo{pages}{1073--1085} (\bibinfo{year}{2015}).

\bibitem{Vietri1997}
\bibinfo{author}{{Vietri}, M.}
\newblock \bibinfo{title}{{GeV Photons from Ultrahigh Energy Cosmic Rays
  Accelerated in Gamma Ray Bursts}}.
\newblock \emph{\bibinfo{journal}{\prl}} \textbf{\bibinfo{volume}{78}},
  \bibinfo{pages}{4328--4331} (\bibinfo{year}{1997}).

\bibitem{Boettcher&Dermer1998}
\bibinfo{author}{{B{\"o}ttcher}, M.} \& \bibinfo{author}{{Dermer}, C.~D.}
\newblock \bibinfo{title}{{High-energy Gamma Rays from Ultra-high-energy
  Cosmic-Ray Protons in Gamma-Ray Bursts}}.
\newblock \emph{\bibinfo{journal}{\apjl}} \textbf{\bibinfo{volume}{499}},
  \bibinfo{pages}{L131--L134} (\bibinfo{year}{1998}).

\bibitem{Gropp2019}
\bibinfo{author}{{Gropp}, J.~D.}
\newblock \bibinfo{title}{{GRB 190114C: Swift detection of a very bright burst
  with a bright optical counterpart}}.
\newblock \emph{\bibinfo{journal}{GRB Coordinates Network}} \textbf{\bibinfo{volume}{23688}}
  (\bibinfo{year}{2019}).

\bibitem{Hamburg2019}
\bibinfo{author}{{Hamburg}, R.}
\newblock \bibinfo{title}{{GRB 190114C: Fermi GBM detection}}.
\newblock \emph{\bibinfo{journal}{GRB Coordinates Network}} \textbf{\bibinfo{volume}{23707}}
  (\bibinfo{year}{2019}).

\bibitem{Krimmetal2019}
\bibinfo{author}{{Krimm}, H.~A.} \emph{et~al.}
\newblock \bibinfo{title}{{GRB 190114C: Swift-BAT refined analysis.}}
\newblock \emph{\bibinfo{journal}{GRB Coordinates Network}}
  \textbf{\bibinfo{volume}{23724}}
  (\bibinfo{year}{2019}).

\bibitem{Selsing2019}
\bibinfo{author}{{Selsing}, J.}
\newblock \bibinfo{title}{{GRB 190114C: NOT optical counterpart and redshift}}.
\newblock \emph{\bibinfo{journal}{GRB Coordinates Network}} \textbf{\bibinfo{volume}{23695}}
  (\bibinfo{year}{2019}).

\bibitem{Castro-Tirado2019}
\bibinfo{author}{{Castro-Tirado}, A.}
\newblock \bibinfo{title}{{GRB 190114C: refined redshift by the 10.4m GTC}}.
\newblock \emph{\bibinfo{journal}{GRB Coordinates Network}} \textbf{\bibinfo{volume}{23708}}
  (\bibinfo{year}{2019}).

\bibitem{Aleksicetal2016a}
\bibinfo{author}{{Aleksi{\'c}}, J.} \emph{et~al.}
\newblock \bibinfo{title}{{The major upgrade of the MAGIC telescopes, Part I:
  The hardware improvements and the commissioning of the system}}.
\newblock \emph{\bibinfo{journal}{Astroparticle Physics}}
  \textbf{\bibinfo{volume}{72}}, \bibinfo{pages}{61--75}
  (\bibinfo{year}{2016}).

\bibitem{Aleksicetal2016b}
\bibinfo{author}{{Aleksi{\'c}}, J.} \emph{et~al.}
\newblock \bibinfo{title}{{The major upgrade of the MAGIC telescopes, Part II:
  A performance study using observations of the Crab Nebula}}.
\newblock \emph{\bibinfo{journal}{Astroparticle Physics}}
  \textbf{\bibinfo{volume}{72}}, \bibinfo{pages}{76--94}
  (\bibinfo{year}{2016}).

\bibitem{Mirzoyan2019}
\bibinfo{author}{{Mirzoyan}, R.}
\newblock \bibinfo{title}{{First time detection of a GRB at sub-TeV energies;
  MAGIC detects the GRB 190114C}}.
\newblock \emph{\bibinfo{journal}{The Astronomer's Telegram}}
  \textbf{\bibinfo{volume}{12390}} (\bibinfo{year}{2019}).

\bibitem{Dwek&Krennrich2013}
\bibinfo{author}{{Dwek}, E.} \& \bibinfo{author}{{Krennrich}, F.}
\newblock \bibinfo{title}{{The extragalactic background light and the gamma-ray
  opacity of the universe}}.
\newblock \emph{\bibinfo{journal}{Astroparticle Physics}}
  \textbf{\bibinfo{volume}{43}}, \bibinfo{pages}{112--133}
  (\bibinfo{year}{2013}).

\bibitem{Dominguezetal11}
\bibinfo{author}{{Dom{\'\i}nguez}, A.} \emph{et~al.}
\newblock \bibinfo{title}{{Extragalactic background light inferred from AEGIS
  galaxy-SED-type fractions}}.
\newblock \emph{\bibinfo{journal}{\mnras}} \textbf{\bibinfo{volume}{410}},
  \bibinfo{pages}{2556--2578} (\bibinfo{year}{2011}).

\bibitem{Ravasioetal2019}
\bibinfo{author}{{Ravasio}, M.~E.} \emph{et~al.}
\newblock \bibinfo{title}{{GRB 190114C: from prompt to afterglow?}}
\newblock \emph{\bibinfo{journal}{\aap}} \textbf{\bibinfo{volume}{626}},
  \bibinfo{pages}{A12} (\bibinfo{year}{2019}).

\bibitem{Wangetal2019}
\bibinfo{author}{{Wang}, X.-Y.}, \bibinfo{author}{{Liu}, R.-Y.},
  \bibinfo{author}{{Zhang}, H.-M.}, \bibinfo{author}{{Xi}, S.-Q.} \&
  \bibinfo{author}{{Zhang}, B.}
\newblock \bibinfo{title}{{Synchrotron self-Compton emission from afterglow
  shocks as the origin of the sub-TeV emission in GRB 180720B and GRB
  190114C}}.
\newblock \emph{\bibinfo{journal}{arXiv e-prints}}
  \bibinfo{pages}{arXiv:1905.11312} (\bibinfo{year}{2019}).

\bibitem{Ackermannetal2013}
\bibinfo{author}{{Ackermann}, M.} \emph{et~al.}
\newblock \bibinfo{title}{{The First Fermi-LAT Gamma-Ray Burst Catalog}}.
\newblock \emph{\bibinfo{journal}{\apjs}} \textbf{\bibinfo{volume}{209}},
  \bibinfo{pages}{11} (\bibinfo{year}{2013}).

\bibitem{Piran&Nakar2010}
\bibinfo{author}{{Piran}, T.} \& \bibinfo{author}{{Nakar}, E.}
\newblock \bibinfo{title}{{On the External Shock Synchrotron Model for
  Gamma-ray Bursts' GeV Emission}}.
\newblock \emph{\bibinfo{journal}{\apjl}} \textbf{\bibinfo{volume}{718}},
  \bibinfo{pages}{L63--L67} (\bibinfo{year}{2010}).

\bibitem{Ackermannetal2014}
\bibinfo{author}{{Ackermann}, M.} \emph{et~al.}
\newblock \bibinfo{title}{{Fermi-LAT Observations of the Gamma-Ray Burst GRB
  130427A}}.
\newblock \emph{\bibinfo{journal}{Science}} \textbf{\bibinfo{volume}{343}},
  \bibinfo{pages}{42--47} (\bibinfo{year}{2014}).

\bibitem{Kouveliotouetal2013}
\bibinfo{author}{{Kouveliotou}, C.} \emph{et~al.}
\newblock \bibinfo{title}{{NuSTAR Observations of GRB 130427A Establish a
  Single Component Synchrotron Afterglow Origin for the Late Optical to
  Multi-GeV Emission}}.
\newblock \emph{\bibinfo{journal}{\apjl}} \textbf{\bibinfo{volume}{779}},
  \bibinfo{pages}{L1} (\bibinfo{year}{2013}).

\setcounter{firstbib}{\value{enumiv}}

\end{thebibliography}

\begin{thebibliography}{10}
\setcounter{enumiv}{\value{firstbib}}
\expandafter\ifx\csname url\endcsname\relax
  \def\url#1{\texttt{#1}}\fi
\expandafter\ifx\csname urlprefix\endcsname\relax\def\urlprefix{URL }\fi
\providecommand{\bibinfo}[2]{#2}
\providecommand{\eprint}[2][]{\url{#2}}

\bibitem{Amatietal2002}
\bibinfo{author}{{Amati}, L.} \emph{et~al.}
\newblock \bibinfo{title}{{Intrinsic spectra and energetics of BeppoSAX
  Gamma-Ray Bursts with known redshifts}}.
\newblock \emph{\bibinfo{journal}{\aap}} \textbf{\bibinfo{volume}{390}},
  \bibinfo{pages}{81--89} (\bibinfo{year}{2002}).

\bibitem{Yonetokuetal2004}
\bibinfo{author}{{Yonetoku}, D.} \emph{et~al.}
\newblock \bibinfo{title}{{Gamma-Ray Burst Formation Rate Inferred from the
  Spectral Peak Energy-Peak Luminosity Relation}}.
\newblock \emph{\bibinfo{journal}{\apj}} \textbf{\bibinfo{volume}{609}},
  \bibinfo{pages}{935--951} (\bibinfo{year}{2004}).

\bibitem{Navaetal2012}
\bibinfo{author}{{Nava}, L.} \emph{et~al.}
\newblock \bibinfo{title}{{A complete sample of bright Swift long gamma-ray
  bursts: testing the spectral-energy correlations}}.
\newblock \emph{\bibinfo{journal}{\mnras}} \textbf{\bibinfo{volume}{421}},
  \bibinfo{pages}{1256--1264} (\bibinfo{year}{2012}).

\bibitem{Vandenbroucke2019}
\bibinfo{author}{{Vandenbroucke}, J.}
\newblock \bibinfo{title}{{GRB 190114C: Search for high-energy neutrinos with
  IceCube}}.
\newblock \emph{\bibinfo{journal}{The Astronomer's Telegram}}
  \textbf{\bibinfo{volume}{12395}} (\bibinfo{year}{2019}).

\bibitem{Barthelmy2016}
\bibinfo{author}{{Barthelmy}, S.}
\newblock \bibinfo{title}{{GCN capabilities and status, and the incorporation
  of LIGO/Virgo}}.
\newblock In \emph{\bibinfo{booktitle}{APS Meeting Abstracts}},
  \bibinfo{pages}{M13.004} (\bibinfo{year}{2016}).

\bibitem{Carosietal2015}
\bibinfo{author}{{Carosi}, A.} \emph{et~al.}
\newblock \bibinfo{title}{{Recent follow-up observations of GRBs in the very
  high energy band with the MAGIC Telescopes}}.
\newblock In \bibinfo{editor}{{Borisov}, A.~S.} \emph{et~al.} (eds.)
  \emph{\bibinfo{booktitle}{34th International Cosmic Ray Conference
  (ICRC2015)}}, vol.~\bibinfo{volume}{34} of
  \emph{\bibinfo{series}{International Cosmic Ray Conference}},
  \bibinfo{pages}{809} (\bibinfo{year}{2015}).

\bibitem{Bertietal2017}
\bibinfo{author}{{Berti}, A.} \& \bibinfo{author}{{MAGIC GRB Group}}.
\newblock \bibinfo{title}{{Search for High Energy emission from GRBs with
  MAGIC}}.
\newblock In \emph{\bibinfo{booktitle}{IAU Symposium}}, vol.
  \bibinfo{volume}{324} of \emph{\bibinfo{series}{IAU Symposium}},
  \bibinfo{pages}{70--73} (\bibinfo{year}{2017}).

\bibitem{Fominetal1994}
\bibinfo{author}{Fomin, V.} \emph{et~al.}
\newblock \bibinfo{title}{{New methods of atmospheric Cherenkov imaging for
  gamma-ray astronomy. I. The false source method}}.
\newblock \emph{\bibinfo{journal}{Astroparticle Physics}}
  \textbf{\bibinfo{volume}{2}}, \bibinfo{pages}{137 -- 150}
  (\bibinfo{year}{1994}).

\bibitem{Ahnen2017}
\bibinfo{author}{{Ahnen}, M.~L.} \emph{et~al.}
\newblock \bibinfo{title}{{Performance of the MAGIC telescopes under
  moonlight}}.
\newblock \emph{\bibinfo{journal}{Astroparticle Physics}}
  \textbf{\bibinfo{volume}{94}}, \bibinfo{pages}{29--41}
  (\bibinfo{year}{2017}).

\bibitem{Frucketal2014}
\bibinfo{author}{{Fruck}, C.} \emph{et~al.}
\newblock \bibinfo{title}{{A novel LIDAR-based Atmospheric Calibration Method
  for Improving the Data Analysis of MAGIC}}.
\newblock \emph{\bibinfo{journal}{ArXiv e-prints}}  (\bibinfo{year}{2014}).

\bibitem{LiMa83}
\bibinfo{author}{{Li}, T.~P.} \& \bibinfo{author}{{Ma}, Y.~Q.}
\newblock \bibinfo{title}{{Analysis methods for results in gamma-ray
  astronomy.}}
\newblock \emph{\bibinfo{journal}{\apj}} \textbf{\bibinfo{volume}{272}},
  \bibinfo{pages}{317--324} (\bibinfo{year}{1983}).

\bibitem{Schmelling1994}
\bibinfo{author}{{Schmelling}, M.}
\newblock \bibinfo{title}{{The method of reduced cross-entropy A general
  approach to unfold probability distributions}}.
\newblock \emph{\bibinfo{journal}{Nuclear Instruments and Methods in Physics
  Research A}} \textbf{\bibinfo{volume}{340}}, \bibinfo{pages}{400--412}
  (\bibinfo{year}{1994}).

\bibitem{Franceschinietal2008}
\bibinfo{author}{{Franceschini}, A.}, \bibinfo{author}{{Rodighiero}, G.} \&
  \bibinfo{author}{{Vaccari}, M.}
\newblock \bibinfo{title}{{Extragalactic optical-infrared background radiation,
  its time evolution and the cosmic photon-photon opacity}}.
\newblock \emph{\bibinfo{journal}{\aap}} \textbf{\bibinfo{volume}{487}},
  \bibinfo{pages}{837--852} (\bibinfo{year}{2008}).

\bibitem{Finkeetal2010}
\bibinfo{author}{{Finke}, J.~D.}, \bibinfo{author}{{Razzaque}, S.} \&
  \bibinfo{author}{{Dermer}, C.~D.}
\newblock \bibinfo{title}{{Modeling the Extragalactic Background Light from
  Stars and Dust}}.
\newblock \emph{\bibinfo{journal}{\apj}} \textbf{\bibinfo{volume}{712}},
  \bibinfo{pages}{238--249} (\bibinfo{year}{2010}).

\bibitem{Gilmoreetal2012}
\bibinfo{author}{{Gilmore}, R.~C.}, \bibinfo{author}{{Somerville}, R.~S.},
  \bibinfo{author}{{Primack}, J.~R.} \& \bibinfo{author}{{Dom{\'\i}nguez}, A.}
\newblock \bibinfo{title}{{Semi-analytic modelling of the extragalactic
  background light and consequences for extragalactic gamma-ray spectra}}.
\newblock \emph{\bibinfo{journal}{\mnras}} \textbf{\bibinfo{volume}{422}},
  \bibinfo{pages}{3189--3207} (\bibinfo{year}{2012}).

\bibitem{XRT_GCN}
\bibinfo{author}{{Beardmore}, A.}
\newblock \bibinfo{title}{{The Swift-XRT WT mode spectrum of GRB190114C.}}
\newblock \emph{\bibinfo{journal}{GRB Coordinates Network}} \textbf{\bibinfo{volume}{23736}}
  (\bibinfo{year}{2019}).

\bibitem{Navaetal2013}
\bibinfo{author}{{Nava}, L.}, \bibinfo{author}{{Sironi}, L.},
  \bibinfo{author}{{Ghisellini}, G.}, \bibinfo{author}{{Celotti}, A.} \&
  \bibinfo{author}{{Ghirlanda}, G.}
\newblock \bibinfo{title}{{Afterglow emission in gamma-ray bursts - I.
  Pair-enriched ambient medium and radiative blast waves}}.
\newblock \emph{\bibinfo{journal}{\mnras}} \textbf{\bibinfo{volume}{433}},
  \bibinfo{pages}{2107--2121} (\bibinfo{year}{2013}).

\bibitem{Totani1998}
\bibinfo{author}{{Totani}, T.}
\newblock \bibinfo{title}{{Very Strong TeV Emission as Gamma-Ray Burst
  Afterglows}}.
\newblock \emph{\bibinfo{journal}{\apj}} \textbf{\bibinfo{volume}{502}},
  \bibinfo{pages}{L13--L16} (\bibinfo{year}{1998}).

\bibitem{Razzaque2010}
\bibinfo{author}{{Razzaque}, S.}
\newblock \bibinfo{title}{{A Leptonic-Hadronic Model for the Afterglow of
  Gamma-ray Burst 090510}}.
\newblock \emph{\bibinfo{journal}{\apj}} \textbf{\bibinfo{volume}{724}},
  \bibinfo{pages}{L109--L112} (\bibinfo{year}{2010}).

\bibitem{Galli&Piro2008}
\bibinfo{author}{{Galli}, A.} \& \bibinfo{author}{{Piro}, L.}
\newblock \bibinfo{title}{{Prospects for detection of very high-energy emission
  from GRB in the context of the external shock model}}.
\newblock \emph{\bibinfo{journal}{\aap}} \textbf{\bibinfo{volume}{489}},
  \bibinfo{pages}{1073--1077} (\bibinfo{year}{2008}).

\bibitem{Connaughtonetal1999}
\bibinfo{author}{{Connaughton}, V.} \& \bibinfo{author}{{VERITAS
  Collaboration}}.
\newblock \bibinfo{title}{{Gamma-ray bursts at VERITAS energies}}.
\newblock \emph{\bibinfo{journal}{Astroparticle Physics}}
  \textbf{\bibinfo{volume}{11}}, \bibinfo{pages}{255--257}
  (\bibinfo{year}{1999}).

\bibitem{Atkinsetal2000}
\bibinfo{author}{{Atkins}, R.} \emph{et~al.}
\newblock \bibinfo{title}{{Evidence for TEV Emission from GRB 970417A}}.
\newblock \emph{\bibinfo{journal}{\apj}} \textbf{\bibinfo{volume}{533}},
  \bibinfo{pages}{L119--L122} (\bibinfo{year}{2000}).

\bibitem{Atkinsetal2004}
\bibinfo{author}{{Atkins}, R.} \emph{et~al.}
\newblock \bibinfo{title}{{Limits on Very High Energy Emission from Gamma-Ray
  Bursts with the Milagro Observatory}}.
\newblock \emph{\bibinfo{journal}{\apjl}} \textbf{\bibinfo{volume}{604}},
  \bibinfo{pages}{L25--L28} (\bibinfo{year}{2004}).

\bibitem{Abdoetal2007}
\bibinfo{author}{{Abdo}, A.~A.} \emph{et~al.}
\newblock \bibinfo{title}{{Milagro Constraints on Very High Energy Emission
  from Short-Duration Gamma-Ray Bursts}}.
\newblock \emph{\bibinfo{journal}{\apj}} \textbf{\bibinfo{volume}{666}},
  \bibinfo{pages}{361--367} (\bibinfo{year}{2007}).

\bibitem{Horanetal2007}
\bibinfo{author}{{Horan}, D.} \emph{et~al.}
\newblock \bibinfo{title}{{Very High Energy Observations of Gamma-Ray Burst
  Locations with the Whipple Telescope}}.
\newblock \emph{\bibinfo{journal}{\apj}} \textbf{\bibinfo{volume}{655}},
  \bibinfo{pages}{396--405} (\bibinfo{year}{2007}).

\bibitem{Aharonianetal2009a}
\bibinfo{author}{{Aharonian}, F.} \emph{et~al.}
\newblock \bibinfo{title}{{HESS Observations of the Prompt and Afterglow Phases
  of GRB 060602B}}.
\newblock \emph{\bibinfo{journal}{\apj}} \textbf{\bibinfo{volume}{690}},
  \bibinfo{pages}{1068--1073} (\bibinfo{year}{2009}).

\bibitem{Aharonianetal2009b}
\bibinfo{author}{{Aharonian}, F.} \emph{et~al.}
\newblock \bibinfo{title}{{HESS observations of {$\gamma$}-ray bursts in
  2003-2007}}.
\newblock \emph{\bibinfo{journal}{\aap}} \textbf{\bibinfo{volume}{495}},
  \bibinfo{pages}{505--512} (\bibinfo{year}{2009}).

\bibitem{Acciarietal2011}
\bibinfo{author}{{Acciari}, V.~A.} \emph{et~al.}
\newblock \bibinfo{title}{{VERITAS Observations of Gamma-Ray Bursts Detected by
  Swift}}.
\newblock \emph{\bibinfo{journal}{\apj}} \textbf{\bibinfo{volume}{743}},
  \bibinfo{pages}{62} (\bibinfo{year}{2011}).

\bibitem{Abramowskietal2014}
\bibinfo{author}{{H.E.S.S.~Collaboration}} \emph{et~al.}
\newblock \bibinfo{title}{{Search for TeV Gamma-ray Emission from GRB 100621A,
  an extremely bright GRB in X-rays, with H.E.S.S.}}
\newblock \emph{\bibinfo{journal}{\aap}} \textbf{\bibinfo{volume}{565}},
  \bibinfo{pages}{A16} (\bibinfo{year}{2014}).

\bibitem{Alfaroetal2017}
\bibinfo{author}{{Alfaro}, R.} \emph{et~al.}
\newblock \bibinfo{title}{{Search for Very-high-energy Emission from Gamma-Ray
  Bursts Using the First 18 Months of Data from the HAWC Gamma-Ray
  Observatory}}.
\newblock \emph{\bibinfo{journal}{\apj}} \textbf{\bibinfo{volume}{843}},
  \bibinfo{pages}{88} (\bibinfo{year}{2017}).

\bibitem{Hoischenetal2017}
\bibinfo{author}{{Hoischen}, C.} \emph{et~al.}
\newblock \bibinfo{title}{{GRB Observations with H.E.S.S. II}}.
\newblock \emph{\bibinfo{journal}{International Cosmic Ray Conference}}
  \textbf{\bibinfo{volume}{35}}, \bibinfo{pages}{636} (\bibinfo{year}{2017}).

\bibitem{Abeysekaraetal2018}
\bibinfo{author}{{Abeysekara}, A.~U.} \emph{et~al.}
\newblock \bibinfo{title}{{A Strong Limit on the Very-high-energy Emission from
  GRB 150323A}}.
\newblock \emph{\bibinfo{journal}{\apj}} \textbf{\bibinfo{volume}{857}},
  \bibinfo{pages}{33} (\bibinfo{year}{2018}).

\bibitem{Albertetal2007}
\bibinfo{author}{{Albert}, J.} \emph{et~al.}
\newblock \bibinfo{title}{{MAGIC Upper Limits on the Very High Energy Emission
  from Gamma-Ray Bursts}}.
\newblock \emph{\bibinfo{journal}{\apj}} \textbf{\bibinfo{volume}{667}},
  \bibinfo{pages}{358--366} (\bibinfo{year}{2007}).

\bibitem{Aleksicetal2010}
\bibinfo{author}{{Aleksi{\'c}}, J.} \emph{et~al.}
\newblock \bibinfo{title}{{MAGIC observation of the GRB 080430 afterglow}}.
\newblock \emph{\bibinfo{journal}{\aap}} \textbf{\bibinfo{volume}{517}},
  \bibinfo{pages}{A5} (\bibinfo{year}{2010}).

\bibitem{Aleksicetal2014}
\bibinfo{author}{{Aleksi{\'c}}, J.} \emph{et~al.}
\newblock \bibinfo{title}{{MAGIC upper limits on the GRB 090102 afterglow}}.
\newblock \emph{\bibinfo{journal}{\mnras}} \textbf{\bibinfo{volume}{437}},
  \bibinfo{pages}{3103--3111} (\bibinfo{year}{2014}).

\bibitem{Ghirlandaetal2016}
\bibinfo{author}{{Ghirlanda}, G.} \emph{et~al.}
\newblock \bibinfo{title}{{Short gamma-ray bursts at the dawn of the
  gravitational wave era}}.
\newblock \emph{\bibinfo{journal}{\aap}} \textbf{\bibinfo{volume}{594}},
  \bibinfo{pages}{A84} (\bibinfo{year}{2016}).

\bibitem{Perleyetal2016}
\bibinfo{author}{{Perley}, D.~A.} \emph{et~al.}
\newblock \bibinfo{title}{{The Swift Gamma-Ray Burst Host Galaxy Legacy Survey.
  I. Sample Selection and Redshift Distribution}}.
\newblock \emph{\bibinfo{journal}{\apj}} \textbf{\bibinfo{volume}{817}},
  \bibinfo{pages}{7} (\bibinfo{year}{2016}).

\bibitem{Lennarzetal2013}
\bibinfo{author}{{Lennarz}, D.} \emph{et~al.}
\newblock \bibinfo{title}{{Searching for TeV emission from GRBs: the status of
  the H.E.S.S. GRB programme}}.
\newblock \emph{\bibinfo{journal}{arXiv e-prints}}
  \bibinfo{pages}{arXiv:1307.6897} (\bibinfo{year}{2013}).

\bibitem{Smithetal2000}
\bibinfo{author}{Smith, A.~J.} \emph{et~al.}
\newblock \bibinfo{title}{Results from the milagrito experiment}.
\newblock \emph{\bibinfo{journal}{AIP Conference Proceedings}}
  \textbf{\bibinfo{volume}{515}}, \bibinfo{pages}{441--447}
  (\bibinfo{year}{2000}).

\bibitem{Aune2009}
\bibinfo{author}{{Aune}, T.}
\newblock \bibinfo{title}{{A Search for GeV-TeV Emission from GRBs Using the
  Milagro Detector}}.
\newblock In \bibinfo{editor}{{Meegan}, C.}, \bibinfo{editor}{{Kouveliotou},
  C.} \& \bibinfo{editor}{{Gehrels}, N.} (eds.)
  \emph{\bibinfo{booktitle}{American Institute of Physics Conference Series}},
  vol. \bibinfo{volume}{1133} of \emph{\bibinfo{series}{American Institute of
  Physics Conference Series}}, \bibinfo{pages}{385--387}
  (\bibinfo{year}{2009}).

\bibitem{SazParkinson2009}
\bibinfo{author}{{Saz Parkinson}, P.~M.}
\newblock \bibinfo{title}{{A search for GeV-TeV emission from Gamma-ray Bursts
  using the Milagro detector}}.
\newblock In \bibinfo{editor}{{Bastieri}, D.} \& \bibinfo{editor}{{Rando}, R.}
  (eds.) \emph{\bibinfo{booktitle}{American Institute of Physics Conference
  Series}}, vol. \bibinfo{volume}{1112}, \bibinfo{pages}{181--186}
  (\bibinfo{year}{2009}).

\bibitem{Evansetal2010}
\bibinfo{author}{{Evans}, P.~A.} \emph{et~al.}
\newblock \bibinfo{title}{{The Swift Burst Analyser. I. BAT and XRT spectral
  and flux evolution of gamma ray bursts}}.
\newblock \emph{\bibinfo{journal}{\aap}} \textbf{\bibinfo{volume}{519}},
  \bibinfo{pages}{A102} (\bibinfo{year}{2010}).

\end{thebibliography}
\end{document}